\documentclass[12pt]{article}

\usepackage{epsfig}
\usepackage{amssymb,amsmath}
\usepackage{multirow}
\usepackage{rotating}
\usepackage{graphicx}
\usepackage{lscape}
\usepackage{hyperref}
\usepackage{xcolor}
\usepackage{url}
\usepackage{feynmp-auto} 

\usepackage{comment}

\newcommand{\bea}{\begin{eqnarray}}
\newcommand{\eea}{\end{eqnarray}}

 \makeatletter
\def\alt{\mathrel{\mathpalette\gl@align<}}
\def\agt{\mathrel{\mathpalette\gl@align>}}
\def\gl@align#1#2{\lower.6ex\vbox{\baselineskip\z@skip\lineskip\z@
\ialign{$\m@th#1\hfil##\hfil$\crcr#2\crcr\sim\crcr}}} \makeatother

\begin{document}

\begin{flushright}
\end{flushright}

\vspace*{1.0cm}

\begin{center}
\baselineskip 20pt 
{\Large\bf 
Muon $g-2$ and semileptonic $B$ decays in 
BDW model
with gauge kinetic mixing
}
\vspace{1cm}

{\large 
Sang Quang Dinh$^{a,}$
and 
Hieu Minh Tran$^{b,}$\footnote{ E-mail: hieu.tranminh@hust.edu.vn} 
} 
\vspace{.5cm}

{\baselineskip 20pt \it
$^a$VNU University of Science, Vietnam National University - Hanoi,  \\
334 Nguyen Trai Road, Hanoi, Vietnam \\
\vspace{2mm} 
$^b$Hanoi University of Science and Technology, 1 Dai Co Viet Road, Hanoi, Vietnam 
}

\vspace{.5cm}

\vspace{1.5cm} {\bf Abstract}
\end{center}
In the model proposed by B\'elanger, Delaunay and Westhoff (BDW),
a new sector consisted of vectorlike fermions and two complex scalars is charged under an extra Abelian symmetry $U(1)_X$.
In this paper, we generalize the BDW model by introducing the kinetic mixing between the $U(1)_X$ and the standard model $U(1)_Y$ gauge fields.
The new physics contributions to the muon anomalous magnetic moment and the Wilson coefficients $C_{9,10}^{(')}$ are obtained analytically.
We have explored the free parameter space of the model, taking into account 
various 
constraints on the muon $g-2$ 
using recent data from the E989 experiment at Fermilab, the lepton universality violation in terms of $R_K$ and $R_{K^*}$,
and the branching ratios of the semileptonic decays, 
$B^+ \rightarrow K^+ \mu^+ \mu^-$ and
$B^0 \rightarrow K^{*0} \mu^+ \mu^-$, 
the LEP and LHC searches for sleptons and $Z'$ boson,
as well as the perturbative requirement.
The viable parameter regions of the model are identified.
In the presence of the gauge kinetic mixing term, those regions are enlarged and significantly deformed in comparison to the case with vanishing kinetic mixing.
In the near future, the E989 experiment 
with the projected sensitivity
will be able to test significant parts of the currently allowed parameter regions.

\thispagestyle{empty}

\newpage

\addtocounter{page}{-1}

\baselineskip 18pt


\section{Introduction}  

Although the standard model (SM) predictions have been verified in many experiments showing an excellent agreement with data, new physics seems to be around the corner due to unanswered questions.
On the one hand, one of the most important precision tests of the SM is the muon anomalous magnetic moment ($g-2$) whose value was determined accurately as
\cite{Tanabashi:2018oca, Abi:2021gix}
\begin{eqnarray}
a_\mu^\text{exp}	&=&
	(11659206.1 \pm 4.1) 
	\times 10^{-10} .
\end{eqnarray}
This is the average value taking into account the new result from the E989 experiment at Fermilab \cite{Abi:2021gix} which confirmed the previous E821 measurement at Brookhaven National Laboratory.
However, the SM prediction for the muon $g-2$ is presently \cite{Aoyama:2020ynm}
\begin{eqnarray}
a_\mu^\text{SM} &=& 
	(11659181.0 \pm 4.3)\times 10^{-10}	\, ,
\end{eqnarray}
corresponding to a 
4.2$\sigma$ deviation from the above world-average experimental value.
According to the projected sensitivity of the E989 experiment
\cite{Grange:2015fou} 
as well as the future experiment (E34) at J-PARC \cite{Mibe:2010zz}, 
the precision will be improved by a factor of four that will shed light on this deviation. 
If the experimental center value of the muon $g-2$ remains unchanged, the above deviation will raise up to about
more than 5$\sigma$
\cite{Keshavarzi:2018mgv}, evidently indicating the existence of new physics coupled to the lepton sector
\cite{Tran:2018kxv}.

On the other hand, 
with the improvement of experimental accuracy, rare decays of $B$-mesons are useful as the precision tests of the SM. The small branching ratios of these processes make them good probes to search for new physics beyond the SM. 
In fact, anomalies have been observed in the rare semileptonic $B$ decays related to the quark transition process $b\rightarrow s \ell^+ \ell^-$. 
In particular, the measured relative branching ratios for $1.1\text{ GeV} < q^2 < 6.0\text{ GeV}$ 
\cite{Aaij:2019wad,Aaij:2017vbb}
\begin{eqnarray}
&& R_K 
	= \frac{BR(B^+ \rightarrow K^+ \mu^+\mu^-)}{BR(B^+ \rightarrow K^+ e^+e^-)} 
	= 0.846^{+0.044}_{-0.041} 
	\, ,	\\
&& R_{K^*} 
	= \frac{BR(B^0 \rightarrow K^{*0} \mu^+\mu^-)}{BR(B^0 \rightarrow K^{*0} e^+e^-)}
	= 0.71^{+0.12}_{-0.09} \, ,
\end{eqnarray}
deviate from the corresponding SM predictions being close to unity \cite{Bouchard:2013mia,Bobeth:2007dw,Hiller:2003js} 
at the levels of 
more than 3$\sigma$ and 2.4$\sigma$, respectively 
\cite{Bordone:2016gaq,Alok:2019ufo}.
This may be a signature of the lepton universality violation  implying the existence of new physics beyond the SM.
Recently, the updated LHCb results on the angular analysis of the decay process $B^0 \rightarrow K^{*0}(892) \mu^+\mu^-$ show a 3.3$\sigma$ deviation from the SM prediction
\cite{Aaij:2020nrf, 
Descotes-Genon:2013vna} that is slightly increased in comparison to the previous observation \cite{Aaij:2015oid}. 
This is due to the higher statistics regarding to the inclusion of 2016 data for 13 TeV collisions.


Model independent analyses on the experimental data using the effective field theory approach with the Hamiltonian \cite{Grinstein:1988me}
\begin{eqnarray}
\mathcal{H}_\text{eff} &=&
	-\frac{\alpha G_F}{\sqrt{2}\pi} V_{tb} V_{ts}^* 
	\sum_i\left( C_i \mathcal{O}_i + C'_i \mathcal{O}'_i  \right) 
	+ h.c. \, ,
\end{eqnarray}
where $\alpha$ and $G_F$ are the fine structure constant and the Fermi constant,
have been performed to determine the new physics contributions to the relevant Wilson coefficients \cite{Alguero:2019ptt}
\begin{eqnarray}
C_i^{(')} &=&
	C_i^{(')\text{SM}} +
	C_i^{(')\text{NP}}	,
\end{eqnarray}
corresponding to the operators
\begin{eqnarray}
\mathcal{O}_7^{(')}	&=&	
	\frac{m_b}{e} 
	\left[\bar{s} \sigma_{\mu\nu} P_{R(L)} b \right] 
	F^{\mu\nu}	\, , \\
\mathcal{O}_9^{(')}	&=&	
	\left[\bar{s} \gamma_{\mu} P_{L(R)} b \right] 
	\left[\bar{\ell} \gamma^\mu \ell \right]	\, , \\
\mathcal{O}_{10}^{(')}	&=&	
	\left[\bar{s} \gamma_{\mu} P_{L(R)} b \right]
	\left[\bar{\ell} \gamma^\mu \gamma_5 \ell \right]	\, , \\
\mathcal{O}_S^{(')}	&=&
	\left[\bar{s} P_{R(L)} b \right]
	\left[\bar{\ell} \ell \right]	\,	,	\\
\mathcal{O}_P^{(')}	&=&
	\left[\bar{s} P_{R(L)} b \right]
	\left[\bar{\ell} \gamma_5 \ell \right]	\,	.
\end{eqnarray}
The Wilson coefficients $C_{S,P}^{(')}$ and $C_7^{(')}$ are strictly constrained by the leptonic decay process $B_s \rightarrow \mu^+\mu^-$ \cite{Alonso:2014csa}
and the radiative $B$ decays \cite{Paul:2016urs}, respectively.
In the case of semileptonic decays of $B$ mesons, the fitting results showed that scenarios with new physics contributions to the Wilson coefficients is much more favored than the pure SM, especially for the coefficient $C_9^\mu$.
While the new physics contributions to muonic Wilson coefficients ($C_9^{\mu(')}, C_{10}^{\mu(')}$) play an important role, the electronic coefficients ($C_9^{e(')}, C_{10}^{e(')}$) turn out to be consistent with the SM predictions.
Three preferable 1D scenarios for the new physics contribution to muonic coefficients have been found to be: 
(\textit{i}) $C_9^\text{NP}$ only, 
(\textit{ii}) $C_9^\text{NP} = -C_{10}^\text{NP}$, and
(\textit{iii}) $C_9^\text{NP} = -C_9^{'\text{NP}}$ 
respectively \cite{Kowalska:2019ley}.
However, the third scenario is disfavored since it predicts $R_{K^*} \approx 1$ \cite{Alok:2017sui}.
The 2D scenarios were also investigated 
\cite{Alok:2019ufo}.
The fitting to the experimental data with complex Wilson coefficients was  performed in Ref. 
\cite{Biswas:2020uaq}.

Many models have been invented to address the anomalies in the semileptonic decays of $B$ mesons, for example those including a $Z'$ boson resulting from an extended gauge symmetry \cite{Sierra:2015fma,Buras:2013dea,Dwivedi:2019uqd}, leptoquarks \cite{Hiller:2014yaa}, 
new physics contributions via loop corrections 
\cite{Gripaios:2015gra},
or supersymmetry \cite{Altmannshofer:2020axr}.
In this paper, we investigate the model proposed by 
B\'elanger, Delaunay and Westhoff (BDW) in Ref. \cite{Belanger:2015nma} taking into account the gauge kinetic mixing. 
In this model, new particles introduced beyond the SM ones are vectorlike quarks and leptons, and two complex scalars. The model's gauge symmetry is $SU(3)_C \times SU(2)_L \times U(1)_Y \times U(1)_X$ where the additional Abelian symmetry is broken spontaneously resulting in a massive gauge boson $Z'$.
In the presence of the gauge kinetic mixing, we calculate analytically the new physics contributions to the muon $g-2$, and the Wilson coefficients $C_{9,10}^{(')}$ which are used to compute the $B$-meson semileptonic decays.
The most updated data from the Heavy Flavor Averaging Group \cite{hflav2018} and the LHCb Collaboration 
\cite{Aaij:2019wad,Aaij:2017vbb}
are used in our consideration to constrain the model's relevant parameters.
Using the projected sensitivity of the muon $g-2$ experiment E989, we study its ability to test the model in the near future.

The structure of the paper is as follows.
In Section 2, we briefly review the structure of the BDW model, then generalize it by introducing the gauge kinetic mixing.
In Section 3, the analytic results of the new physics contributions to the muon $g-2$ and the Wilson coefficients calculations are presented.
In Section 4, the phenomenological constraints are used to specify the allowed parameter space.
Finally, Section 5 is devoted to conclusions.

\section{BDW model with kinetic mixing}

\subsection{The model}

The new particles introduced in the BDW model beside the SM particles are the vectorlike lepton and quark doublets of the gauge group $SU(2)_L$,
\begin{eqnarray}
L_{L,R} = 
	\begin{pmatrix}
	N_{L,R} \\ E_{L,R}
	\end{pmatrix}	,	\qquad
Q_{L,R} = 
	\begin{pmatrix}
	U_{L,R} \\ D_{L,R}
	\end{pmatrix}	,
\end{eqnarray}
and two complex scalars, $\chi$ and $\phi$, that are singlets under the SM gauge groups.
The symmetry of this model is an extension of the SM symmetry by adding an extra Abelian gauge group, namely 
$SU(3)_C \otimes SU(2)_L \otimes U(1)_Y \otimes U(1)_X$.
The SM particles are invariant under $U(1)_X$ transformation, while the new particles transform nontrivially with the $U(1)_X$ charges given in Table \ref{NP} together with other properties.


\begin{table}[h]
\caption{Properties of new particles introduced in the model \cite{Belanger:2015nma}.}
\label{NP}
\begin{center}
\begin{tabular}{|c|c|c|c|c|c|}
\hline
Particles	&	Spin	&	$SU(3)_C$	&	$SU(2)_L$	&	$U(1)_Y$	&	$U(1)_X$	\\
\hline
$L_L, L_R$	&	1/2	&	\textbf{1}	&	\textbf{2}	&	-1/2	&	1	\\
$Q_L, Q_R$	&	1/2	&	\textbf{3}	&	\textbf{2}	&	1/6	&	-2	\\
\hline
$\chi$		&	0	&	\textbf{1}	&	\textbf{1}	&	0	&	-1	\\
$\phi$		&	0	&	\textbf{1}	&	\textbf{1}	&	0	&	2	\\
\hline
\end{tabular}
\end{center}
\end{table}

The Lagrangian include the SM part and the part involving new physics:
\begin{eqnarray}
\mathcal{L} &=& \mathcal{L}_\text{SM} + \mathcal{L}_\text{NP},
\end{eqnarray}
where
\begin{eqnarray}
\mathcal{L}_\text{NP} & \supset &
	- \; \lambda_{\phi H} |\phi|^2 |H|^2
	- \lambda_{\chi H} |\chi|^2 |H|^2
	- \left[
	y \overline{\ell_L} L_R \chi +
	w \overline{q_L} Q_R \phi + h.c.
	\right] 
	- V_0(\phi,\chi) \nonumber	\\
&&
	- \; ( M_L \overline{L_L} L_R
	+ M_Q \overline{Q_L} Q_R	+ h.c. ) \; ,
\label{Lnp}
\end{eqnarray}
Here, the SM left-handed lepton and quark doublets are denoted as
\begin{eqnarray}
\ell_L^i = 
	\begin{pmatrix}
	\nu^e_L \\ e_L
	\end{pmatrix}_i ,	\qquad
q_L^i =
	\begin{pmatrix}
	u_L	\\ d_L
	\end{pmatrix}_i ,	\qquad
(i = 1,2,3) .
\end{eqnarray}
The explicit form of the scalar potential $V_0(\chi,\phi)$ is given by
\begin{eqnarray}
V_0(\chi,\phi) &=&
	\lambda_\phi |\phi|^4 + m^2_\phi |\phi|^2 +
	\lambda_\chi |\chi|^4 + m^2_\chi |\chi|^2 + 
	\lambda_{\phi\chi} |\phi|^2 |\chi|^2 +
	\left( r \phi \chi^2 + h.c. \right) .
\label{V0}
\end{eqnarray}

Among the new scalars, we assume that only $\phi$ can develop a vacuum expectation value (VEV),
\begin{eqnarray}
\langle \phi \rangle &=& 		
	\sqrt{\frac{-m'^2_\phi}	
		{2\lambda_\phi}}	,
\end{eqnarray}
where
\begin{eqnarray}
m'^2_\phi &=& m^2_\phi + \lambda_{\phi H} \langle H \rangle^2	,
\end{eqnarray}
with $\langle H \rangle = 174$ GeV being the VEV of the SM Higgs field.
Due to the nonzero VEV, $\langle \phi \rangle$, the gauge group $U(1)_X$ is spontaneously broken, leading to a massive $Z'$ boson with a mass
\begin{eqnarray}
m_{Z'} &=& 2\sqrt{2} g_X \langle \phi \rangle	.
\label{mZp}
\end{eqnarray}
where $g_X$ is the $U(1)_X$ gauge coupling.

By decomposing the complex scalar field $\phi$ into their real and imaginary components,
\begin{eqnarray}
\phi &=& 
	\langle \phi \rangle + 
	\frac{1}{\sqrt{2}}
	\left( \varphi_r + i \varphi_i \right)	,
\label{phi}
\end{eqnarray} 
the masses of these fields are found to be
\begin{eqnarray}
m_{\varphi_r} &=& 
	2 \sqrt{\lambda_\phi} \langle \phi \rangle,	\\
m_{\varphi_i} &=& 	0	\,	,
\end{eqnarray}
respectively.
Note that $\varphi_i$ is a massless Nambu-Goldstone boson that can be absorbed by $Z'$ in the unitary gauge.
For the case of $\chi$, after the decomposition
\begin{eqnarray}
\chi &=& 
	\frac{1}{\sqrt{2}}
	\left( \chi_r + i \chi_i \right)	\, ,
\end{eqnarray} 
the mass matrix for these real scalar fields is obtained:
\begin{eqnarray}
\frac{1}{2}
	\begin{pmatrix}
	\chi_r	&	\chi_i
	\end{pmatrix}
M^2_\chi
	\begin{pmatrix}
	\chi_r	\\	\chi_i
	\end{pmatrix}
&=&
\frac{1}{2}
	\begin{pmatrix}
	\chi_r	&	\chi_i
	\end{pmatrix}
	\begin{pmatrix}
	m'^2_\chi + (r+r^*) \langle \phi \rangle	&	i(r-r^*) \langle \phi \rangle	\\
	i(r-r^*) \langle \phi \rangle	&	m'^2_\chi - (r+r^*) \langle \phi \rangle
	\end{pmatrix}
	\begin{pmatrix}
	\chi_r	\\	\chi_i
	\end{pmatrix}	,
\end{eqnarray}
where
\begin{eqnarray}
m'^2_\chi	&=&
	m^2_\chi + 
	\lambda_{\chi H} \langle H \rangle^2 +
	\lambda_{\phi\chi} \langle \phi \rangle^2	.
\end{eqnarray}
In the simple case where the coupling $r$ is real, the matrix $M_\chi^2$ is diagonal, and the masses of the particles $\chi_r$ and $\chi_i$ are respectively
\begin{eqnarray}
m_{\chi_r}	&=&	m'^2_\chi + 2r \langle \phi \rangle	,	\\
m_{\chi_i}	&=&	m'^2_\chi - 2r \langle \phi \rangle	.
\end{eqnarray}


Since the field $\chi$ does not develop a nonzero VEV, there is no mass mixing between the SM leptons and the vectorlike ones.
However, the situation for quarks is more involved because  the VEV of $\phi$ generates mass mixing terms via the new Yukawa interactions with the couplings 
$w = ( w_1, w_2, w_3 )$ in Eq. (\ref{Lnp}).
To diagonalize the quark mass matrices, $M^u$ and $M^d$, we need to use four $4\times 4$ unitary matrices to transform the quark gauge eigenstates, 
$(u^1, u^2, u^3, U)$ and 
$(d^1, d^2, d^3, D)$,
into the mass eigenstates, 
$(u, c, t, \mathcal{U})$ and 
$(d, s, b, \mathcal{D})$:
\begin{eqnarray}
\begin{pmatrix}
u_{L,R} \\	c_{L,R} \\	t_{L,R} \\	\mathcal{U}_{L,R}
\end{pmatrix}	=
	\left( V^u_{L,R} \right)_{4 \times 4}
	\begin{pmatrix}
	u_{L,R}^1 \\	u_{L,R}^2 \\
	u_{L,R}^3 \\	U_{L,R}	
	\end{pmatrix}	,	\qquad
\begin{pmatrix}
d_{L,R} \\	s_{L,R} \\	b_{L,R} \\	\mathcal{D}_{L,R}
\end{pmatrix}	=
	\left( V^d_{L,R} \right)_{4 \times 4}
	\begin{pmatrix}
	d_{L,R}^1 \\	d_{L,R}^2 \\
	d_{L,R}^3 \\	D_{L,R}	
	\end{pmatrix}	.
\label{qmix}
\end{eqnarray}
The diagonal mass matrices of up-type and down-type quarks then read
\begin{eqnarray}
M^u_\text{diag}	&=&	V^u_L M^u (V^u_R)^\dagger	,	\\
M^u_\text{diag}	&=&	V^d_L M^d (V^d_R)^\dagger	.
\end{eqnarray}

\subsection{The gauge kinetic mixing}

In a model with two Abelian gauge symmetries, the gauge kinetic mixing term  of the form $-\frac{k}{2} F^1_{\mu\nu} F^{2\mu\nu}$
can be introduced in the Lagrangian without violating any symmetry \cite{Holdom:1985ag}.
Here, $F^1_{\mu\nu}$, $F^{2\mu\nu}$ and $k$ are respectively the field strength tensors of the $U(1)_Y$ and $U(1)_X$ gauge fields and the kinetic mixing coefficient. 
In fact, the gauge kinetic mixing term is always generated radiatively
\cite{delAguila:1988jz},
even though it is set to zero at high energy scales \cite{Fonseca:2011vn}.
In the presence of such term, the gauge kinetic part of the Lagrangian relating to the Abelian groups can be written as
\begin{eqnarray}
\mathcal{L}^\text{gauge}_\text{kinetic} & \supset &
	-\frac{1}{4}
	\left(
		\begin{matrix}
		F^1_{\mu\nu}	&	F^2_{\mu\nu}		
		\end{matrix}
	\right)
	\left(
		\begin{matrix}
		1	&	k	\\
		k	&	1
		\end{matrix}
	\right)
	\left(
		\begin{matrix}
		F^{1\mu\nu}	\\	F^{2\mu\nu}		
		\end{matrix}
	\right) .
\label{Lkinetic}
\end{eqnarray}
By an appropriate transformation in the space of the Abelian gauge fields, the kinetic Lagrangian can be made canonical.
In the new basis, the covariant derivative is then expressed as
\begin{eqnarray}
D_\mu 
&\supset&	
	\partial_\mu - i Y g_Y B_\mu - i X' g'_X \mathcal{Z}'_\mu \, ,
\end{eqnarray}
where the new charge $X'$ and the new gauge coupling $g'_X$ are determined by the original quantities and the kinetic mixing coefficient $k$ as
\begin{eqnarray}
X' &=&  
	\frac{-k g_Y}{g_X} Y + 
	X ,\\
g'_X	&=&	\frac{g_X}{\sqrt{1-k^2}} .
\end{eqnarray}
Here, $Y(X)$ and $g_{Y}(g_X)$ are the charge and the gauge coupling corresponding to the Abelian group $U(1)_{Y}(U(1)_{X})$, respectively.
The nonzero kinetic mixing coefficient $k$ induces a shift in the $U(1)_X$ charge and modifies the relevant gauge coupling.

After the electroweak symmetry breaking by the VEV of the SM Higgs field, $\langle H \rangle$, the kinetic coefficient generates a mass mixing between the $\mathcal{Z}$ and $\mathcal{Z}'$ bosons, leading to a non-diagonal mass matrix for these particles:
\begin{eqnarray}
M^2_{\mathcal{ZZ}'}	&=&
	\left(	
	\begin{matrix}
	\left( g_2^2 + g_Y^2 \right)\dfrac{\langle H \rangle^2}{2}	&	
	-X'_H g'_X 	\sqrt{g_2^2 + g_Y^2} 
\langle H \rangle^2	\\
	-X'_H g'_X \sqrt{g_2^2 + g_Y^2} \langle H \rangle^2	&	
	2{g'_X}^2 \left( {X'_H}^2 \langle H \rangle^2 + {X'_\phi}^{2} \langle \phi \rangle^2 \right) 
	\end{matrix}
	\right) .
\end{eqnarray}
To diagonalize the above mass matrix, we use the following orthogonal rotation:
\begin{eqnarray}
\left(
\begin{matrix}
\mathcal{Z}_\mu \\	\mathcal{Z}'_\mu
\end{matrix}
\right)	&=&
	\left(
	\begin{matrix}
	\cos \alpha_Z	&	-\sin \alpha_Z	\\
	\sin \alpha_Z	&	\cos \alpha_Z
	\end{matrix}
	\right)
	\left(
	\begin{matrix}
	Z_\mu \\	Z'_\mu
	\end{matrix}
	\right) ,
\end{eqnarray}
where $Z$ and $Z'$ are the mass eigenstates, and the mixing angle $\alpha_Z$ is determined as
\begin{eqnarray}
\tan 2\alpha_Z	&=&
	\frac{
		2\left(M_{\mathcal{ZZ}'}^2 \right)_{12}}	
		{\left(M_{\mathcal{ZZ}'}^2\right)_{11} -
		 \left(M_{\mathcal{ZZ}'}^2\right)_{22}}	\, .
\label{alphaZ}
\end{eqnarray}
It is worth noticing that, in the limit of no kinetic mixing ($k = 0$), 
the pure BDW model is recovered, namely $X'=X$, $g'_X = g_X$, and $\alpha_Z =0$.



From the covariant derivative of muon, we find the interaction terms between the muon and the $Z$ and $Z'$ bosons as
\begin{eqnarray}
\mathcal{L}	&\supset&
	\bar{\mu}
	\left[
		\left(
		g_V \cos \alpha_Z + g_V^k \sin \alpha_Z
		\right) \gamma^\mu +
		\left(
		g_A \cos \alpha_Z + g_A^k \sin \alpha_Z
		\right) \gamma^\mu \gamma^5
	\right] \mu Z_\mu +	\nonumber	\\
&&
	\bar{\mu} 
	\left[
		\left(
		-g_V \sin \alpha_Z + g_V^k \cos \alpha_Z
		\right) \gamma^\mu +
		\left(
		-g_A \sin \alpha_Z + g_A^k \cos \alpha_Z
		\right) \gamma^\mu \gamma^5
	\right] \mu Z'_\mu	\, ,
\end{eqnarray}
where
\begin{eqnarray}
g_V		&=&	
	\frac{g_2}{\cos \theta_W}		
	\left(
		-\frac{1}{4} + \sin^2 \theta_W
	\right)	,	\\
g_A		&=&	
	\frac{g_2}{4\cos \theta_W}	\, ,	\\
X'_{\mu_L}	&=&	
	\frac{-k g_Y}{g_X} Y_{\mu_L}
	\; = \;
	\frac{k g_Y}{2g_X}	\, ,	\\
X'_{\mu_R}	&=&
	\frac{-k g_Y}{g_X} Y_{\mu_R}
	\; = \;
	\frac{k g_Y}{g_X}	\, ,	\\
g_V^k	&=&	
	g'_X \frac{X'_{\mu_L} + X'_{\mu_R}}{2}	
	\; = \;
	\frac{3 k g_Y}{4\sqrt{1-k^2}}	\,	,	\\
g_A^k	&=&	
	g'_X \frac{-X'_{\mu_L} + X'_{\mu_R}}{2}
	\; = \;
	\frac{k g_Y}{4\sqrt{1-k^2}}	\,	.
\end{eqnarray}
In the limit of no kinetic mixing, muons interact only with the $Z$ boson just as in the SM.
Similarly, the interaction terms between the new charged lepton $E_R$ and these two massive neutral gauge bosons are obtained as
\begin{eqnarray}
\mathcal{L}	&\supset &
	\left[
		\frac{g_2 \cos \alpha_Z}{\cos \theta_W} \left( -\frac{1}{2} + \sin^2 \theta_W \right) +
		g'_X X'_{L_R} \sin \alpha_Z
	\right]
	\overline{E_R} \gamma^\mu E_R Z_\mu \nonumber	\\
&&	
	+ \; 
	\left[
		-\frac{g_2 \sin \alpha_Z}{\cos \theta_W} \left( -\frac{1}{2} + \sin^2 \theta_W \right) +
		g'_X X'_{L_R} \cos \alpha_Z
	\right]
	\overline{E_R} \gamma^\mu E_R Z'_\mu \, ,
\end{eqnarray}
where 
\begin{eqnarray}
X'_{L_R} &=& 
	\frac{-k g_Y}{g_X} Y_{L_R} + X_{L_R}	
\; = \;
	\frac{k g_Y}{2 g_X} + 1  .
\end{eqnarray}
For the new scalar sector, the derivation of $Z'$ and $Z$ couplings with $\chi_{r,i}$ from their covariant derivatives leads to
\begin{eqnarray}
\mathcal{L}	&\supset&
	g'_X X'_\chi 
	\sin \alpha_Z Z^\mu 
		\left(
		\partial_\mu \chi_r \cdot \chi_i -
		\chi_r \cdot \partial_\mu \chi_i
		\right) +
	g'_X X'_\chi 
	\cos \alpha_Z Z'^\mu 
		\left(
		\partial_\mu \chi_r \cdot \chi_i -
		\chi_r \cdot \partial_\mu \chi_i
		\right) ,
\end{eqnarray}
where $X'_\chi = X_\chi =-1$.

The flavor changing neutral currents (FCNCs) in the quark sector are induced at the tree level due to the mixing between the SM quarks and the vectorlike quarks.
We parameterize the couplings between the $b$-quark, the $s$-quark and the massive neutral gauge bosons as follows
\begin{eqnarray}
\mathcal{L}	&\supset &
	X'_Q g'_X 
	\left( 
		A_{bs} \overline{b_L} \gamma^\mu s_L +
		B_{bs} \overline{b_R} \gamma^\mu s_R
	\right) \mathcal{Z}'_\mu + 
	\frac{g_2}{\cos \theta_W} C_{bs} 
	\overline{b_R} \gamma^\mu s_R \mathcal{Z}_\mu + h.c. \nonumber	\\
&=&
	\left[
		\left(
		X'_Q g'_X A_{bs} \sin \alpha_Z
		\right) \overline{b_L} \gamma^\mu s_L +
		\left(
		X'_Q g'_X B_{bs} \sin \alpha_Z +
		\frac{g_2}{\cos \theta_W} C_{bs} \cos \alpha_Z
		\right) \overline{b_R} \gamma^\mu s_R
	\right] Z_\mu + \nonumber	\\
&&
	\left[
		\left(
		X'_Q g'_X A_{bs} \cos \alpha_Z
		\right) \overline{b_L} \gamma^\mu s_L +
		\left(
		X'_Q g'_X B_{bs} \cos \alpha_Z -
		\frac{g_2}{\cos \theta_W} C_{bs} \sin \alpha_Z
		\right) \overline{b_R} \gamma^\mu s_R
	\right] Z'_\mu
	+ h.c.
	\, ,	\nonumber	\\
\end{eqnarray}
where 
\begin{eqnarray}
X'_Q &=& -2 - \frac{k g_Y}{6 g_X} \,	,	
\end{eqnarray}
and the parameters $A_{bs}, B_{bs}$ and $C_{bs}$ characterize the FCNCs in the $b \rightarrow s$ transition.
At the tree level, these parameters are determined as
\begin{eqnarray}
A_{bs}	&=&	
	\left(
	V^d_L \cdot \text{Diag}(0,0,0,1) \cdot V_L^{d\dagger}
	\right)_{32}
	\;=\;	
	\left( V^d_L \right)_{34}
	\left( V_L^{d} \right)_{24}^*	\, ,	\\
B_{bs}	&=&
	\left(
	V^d_R \cdot \text{Diag}(0,0,0,1) \cdot V_R^{d\dagger}
	\right)_{32}	
	\;=\;
	\left( V^d_R \right)_{34}
	\left( V_R^{d} \right)_{24}^*	\,	, \\
C_{bs}	&=&
	\left( 
	V^d_R \cdot 
	\text{Diag}
		\left(
		\tfrac{1}{3}\sin^2 \theta_W ,
		\tfrac{1}{3}\sin^2 \theta_W ,
		\tfrac{1}{3}\sin^2 \theta_W ,
		-\tfrac{1}{2} + \tfrac{1}{3}\sin^2 \theta_W
		\right) 
	\cdot V_R^{d\dagger} \right)_{32} 	\, .
\end{eqnarray}
Assuming that the mixing between the vectorlike quarks and the first generation quarks is negligible for simplicity, the parameter $C_{bs}$ can be approximated as
\begin{eqnarray}
C_{bs} & \approx &
	\left(	
	-\frac{1}{2} + \frac{1}{3}\sin^2 \theta_W
	\right)
	B_{bs}	\,	.
\end{eqnarray} 
At the loop-level, these parameters are considered to be the effective couplings encoding the new physics relevant to the quark sector.

\section{New physics contributions to muon $g-2$ and Wilson coefficients}

\subsection{Muon $g-2$}

\begin{figure}[h]
\begin{center}
\includegraphics[scale=1.1]{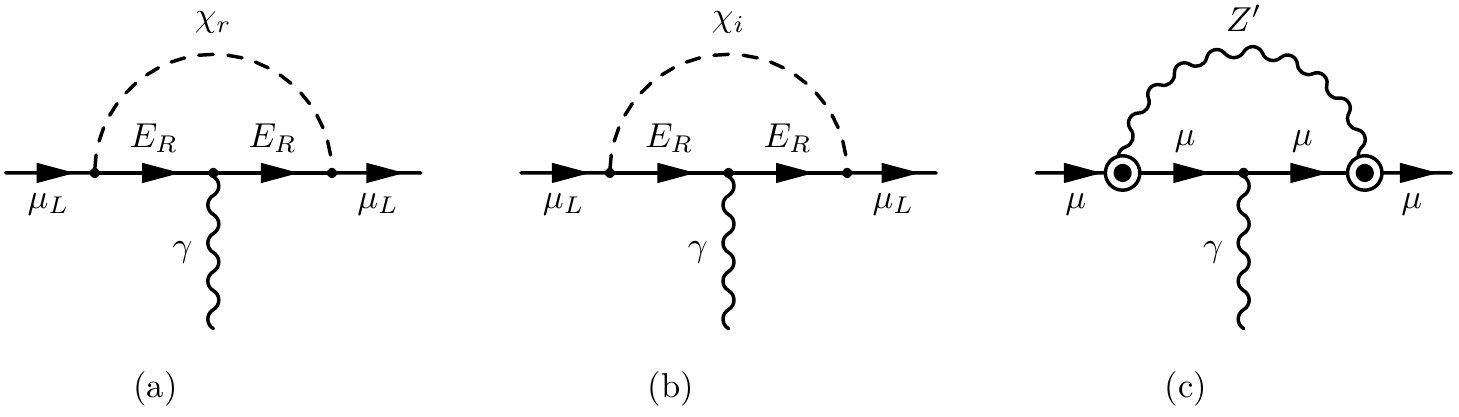}
	\caption{Leading new physics contributions to the muon $g-2$. 
The vertices purely induced by the kinetic mixing effect are represented by an empty circle surrounding a dot.}	
\label{Feyn_g-2}
\end{center}
\end{figure}

In this model, new physics contributes to the muon anomalous magnetic moment via the gauge interaction associated with the massive boson $Z'$, and the new Yukawa interaction between the scalar $\chi$, the right-handed lepton $E_R$ and the muon.
The Feynman diagrams corresponding to the leading new physics contributions to the muon $g-2$ are shown in Figure \ref{Feyn_g-2}.
The diagrams (a) and (b) in this figure are due to the Yukawa coupling $y_\mu$ in Eq. (\ref{Lnp}).
The contribution related to the diagram (c) is generated from the gauge kinetic mixing effect.

From the matrix elements of these one-loop diagrams, after performing some algebraic calculation, we obtain the following expression for the new physics contributions to the muon $g-2$:

\begin{eqnarray}
\Delta a_\mu^\text{NP}	&=&
	\frac{|y_\mu|^2 m_\mu^2}{32\pi^2 m_{\chi_r}^2}
	\left[
		F_g(\tau) + 
		\left( \frac{1}{1+\delta}	\right)
		F_g \left( \frac{\tau}{1+\delta} \right)
	\right] \nonumber	\\
&&
	+\; \frac{\beta}{4\pi^2}
	\int_0^1 dz (1-z)
	\left\{ \vphantom{\frac{1}{1}}
	\left( - g_A \sin \alpha_Z + g_A^k \cos \alpha_Z \right)^2
	(3z-1) 
	\ln \left[ \beta(1-z)^2 +z \right]	
	\right.	\nonumber	\\
&&
	\left.	\qquad
	+ \; \frac{
		\left( - g_V \sin \alpha_Z + g_V^k \cos \alpha_Z  \right)^2 z(1-z) - 
		\left( -g_A \sin \alpha_Z + g_A^k \cos \alpha_Z  \right)^2 z(z+3)	}{\beta(1-z)^2+z}
	\right\}	 , \nonumber	\\
\label{amu_k}
\end{eqnarray}
where
\begin{eqnarray}
\tau	&=&	
	\frac{m_L^2}{m_{\chi_r}^2} ,	\\
\delta	&=&
	\frac{m_{\chi_i}^2}{m_{\chi_r}^2} - 1	,	
	\label{delta_define}	
	\\
\beta	&=&
	\frac{m_\mu^2}{m_{Z'}^2} .
\end{eqnarray}
The first term in Eq. (\ref{amu_k}) with the squared brackets corresponds to the new physics contributions in Figures \ref{Feyn_g-2}a and \ref{Feyn_g-2}b.
Here, the loop function $F_g(x)$ is defined as
\begin{eqnarray}
F_g (x)	&=&
	\frac{1}{6(1-x)^4}
	\left(
		6x \ln x + x^3 -6x^2 + 3 x +2
	\right)	.
\end{eqnarray}
This term is in agreement with the result in Ref. \cite{Belanger:2015nma} for the case of no kinetic mixing 
The second term in Eq. (\ref{amu_k}) with the integral results from the effect of the gauge kinetic mixing via the diagram in Figure \ref{Feyn_g-2}c.
It vanishes in the limit $k=0$.
Since the second term is suppressed by the factor $\beta$ as well as the kinetic mixing coefficient, the sign of $\Delta a_\mu^{NP}$ is determined by the sign of the first term in Eq. (\ref{amu_k}) that is always positive.

\subsection{Wilson coefficients}


\begin{figure}[h!]
\begin{center}
\includegraphics[scale=1.]{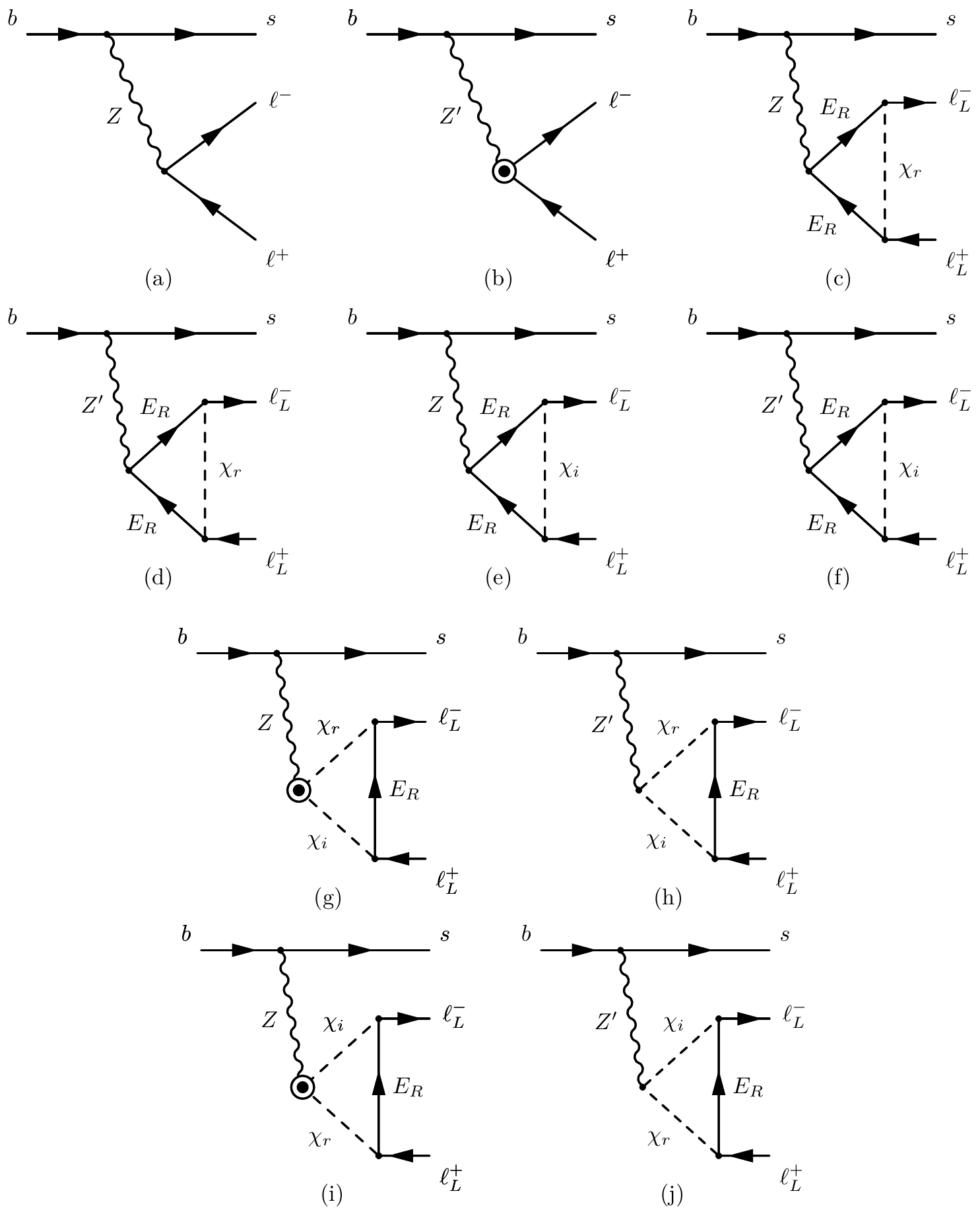}
\caption{Leading new physics contributions to the Wilson coefficients $C_9^{(')}$ and $C_{10}^{(')}$.
The vertices purely induced by the kinetic mixing effect are represented by an empty circle surrounding a dot.} 
\label{Feyn_bsmm}
\end{center}
\end{figure}


Since the scalar and pseudoscalar Wilson coefficients ($C_{S,P}$) are severely restricted by the leptonic decay $B_s \rightarrow \mu^+ \mu^-$ \cite{Alonso:2014csa}, in this subsection we consider the Wilson coefficients $C_{9,10}^{(')}$ to be used for the calculation of the semileptonic branching ratios of $B$ mesons.
The leading new physics contributions to the Wilson coefficients $C_9^{(')}$ and $C_{10}^{(')}$ are depicted by the Feynman diagrams in Figure \ref{Feyn_bsmm}.
Here, we consider the diagrams whose the matrix elements are proportional to the second order of the gauge couplings $g_2$ or $g_X$.

In Figure \ref{Feyn_bsmm}a, the new physics enters this tree-level diagram only via the coupling of the $bsZ$ vertex resulting from the mixing between the SM quarks and the vectorlike quarks.
The contribution according to the diagram \ref{Feyn_bsmm}b is due to the gauge kinetic mixing effect.
The contributions due to the diagrams (c)-(f) stem from the gauge interactions of the vectorlike charged lepton $E_R$ with both $Z$ and $Z'$ bosons.
The cases for the diagrams (g)-(j) relevant to the gauge interactions of the scalars $\chi_{r,i}$ are more involved.
While the contributions from the diagrams (h) and (j) always exist, those from the diagrams (g) and (i) are due to the $ZZ'$ mixing that only emerges when $k \neq 0$.





The new physics contributions to the Wilson coefficients $C_{9,10}^{(')}$ according to the diagrams in Figure \ref{Feyn_bsmm} are expressed as
\begin{eqnarray}
C^{\text{NP}}_9	&=&	
	\frac{12 g_X + k g_Y}{12 (1-k^2)}
	\left(
		\frac{g_X \cos \alpha_Z}{ m_{Z'}^2} A_k +
		\frac{g_2 \sin \alpha_Z}{ m_Z^2} A_k^Z
	\right)
	\Lambda_\text{SM}^2
	\frac{|V_{tb} V_{ts}^*|}{V_{tb} V_{ts}^*} A_{bs}	,	\\
C'^{\text{NP}}_9	&=&	
	\left\{
		\frac{12 g_X + k g_Y}{12(1-k^2) }
	\left(
		\frac{g_X \cos \alpha_Z}{m_{Z'}^2} A_k +
		\frac{g_2 \sin \alpha_Z}{m_Z^2} A_k^Z
	\right)	
	\right.		\nonumber	\\
&&
	\left.	 +
		\frac{g_2 \left( 
		-\frac{1}{2} + \frac{1}{3} \sin^2 \theta_W
		\right)}{2\cos \theta_W \sqrt{1-k^2}}
		\left[
		\frac{g_X\sin \alpha_Z}{m_{Z'}^2} A_k -
		\frac{g_2 \cos \alpha_Z}{m_Z^2} A_k^Z
		\right]
	\right\} 
	\Lambda_\text{SM}^2
	\frac{|V_{tb} V_{ts}^*|}{V_{tb} V_{ts}^*} B_{bs}	,	 
	\\
C^{\text{NP}}_{10}	&=&	
	\frac{12 g_X + k g_Y}{12 (1-k^2)}
	\left(
		\frac{g_X \cos \alpha_Z}{ m_{Z'}^2} B_k +
		\frac{g_2 \sin \alpha_Z}{ m_Z^2} B_k^Z
	\right)
	\Lambda_\text{SM}^2
	\frac{|V_{tb} V_{ts}^*|}{V_{tb} V_{ts}^*} A_{bs}	,\\
C'^{\text{NP}}_{10}	&=&
	\left\{
		\frac{12 g_X + k g_Y}{12(1-k^2) }
	\left(
		\frac{g_X \cos \alpha_Z}{m_{Z'}^2} B_k +
		\frac{g_2 \sin \alpha_Z}{m_Z^2} B_k^Z
	\right)	
	\right.		\nonumber	\\
&&
	\left.	 +
		\frac{g_2 \left( 
		-\frac{1}{2} + \frac{1}{3} \sin^2 \theta_W
		\right)}{2\cos \theta_W \sqrt{1-k^2}}
		\left[
		\frac{g_X\sin \alpha_Z}{m_{Z'}^2} B_k -
		\frac{g_2 \cos \alpha_Z}{m_Z^2} B_k^Z
		\right]
	\right\} 
	\Lambda_\text{SM}^2
	\frac{|V_{tb} V_{ts}^*|}{V_{tb} V_{ts}^*} B_{bs}	,
\end{eqnarray}
where the intermediate notations $A_k$, $A^Z_k$, $B_k$, and $B^Z_k$ are defined as follows
\begin{eqnarray}
A_k	(q^2)	&=&	
	\left[
		\frac{3k g_Y \cos \alpha_Z}{4g_X} - 
		\frac{g_2 \sqrt{1-k^2} \sin \alpha_Z}{g_X \cos \theta_W}
		\left(-\frac{1}{4} + \sin^2 \theta_W	\right)
	\right]	
	+ \, \frac{|y_\ell|^2 \cos \alpha_Z}{32\pi^2}	f_A 
	\nonumber	\\
&&
	+ \, \frac{|y_\ell|^2}{32\pi^2}
	\left[
		\left( 1+ \frac{kg_Y}{2g_X} \right) \cos \alpha_Z -
		\frac{g_2 \sqrt{1-k^2} \sin \alpha_Z}{g_X \cos \theta_W}
		\left( -\frac{1}{2} + \sin^2 \theta_W \right)
	\right] g_A	\, ,
\end{eqnarray}

\begin{eqnarray}
A^Z_k (q^2)	&=&		
	\left[
	\frac{\sqrt{1-k^2}}{\cos \theta_W} 
	\left( -\frac{1}{4} + \sin^2 \theta_W \right)
	\cos \alpha_Z +
	\frac{3k g_Y}{4g_2} \sin \alpha_Z
	\right]	
	+ \, \frac{|y_\ell|^2 g_X \sin \alpha_Z}{32\pi^2 g_2} f_A
	\nonumber	\\
&&	
	+ \, \frac{|y_\ell|^2}{32\pi^2}
	\left[
	\frac{\sqrt{1-k^2} \cos \alpha_Z}{\cos \theta_W}
	\left( -\frac{1}{2} + \sin^2 \theta_W \right) +
	\left( 1 + \frac{k g_Y}{2 g_X} \right)
	\frac{g_X \sin \alpha_Z}{g_2}
	\right]	g_A \, ,
\end{eqnarray}
\begin{eqnarray}
B_k	(q^2)	&=&	
	\left[
	\frac{k g_Y \cos \alpha_Z}{4g_X} -
	\frac{g_2 \sqrt{1-k^2} \sin \alpha_Z}{4g_X \cos \theta_W}
	\right] 
	- \, \frac{|y_\ell|^2 \cos \alpha_Z}{32\pi^2} f_B
	\nonumber	\\
&&
	+ \, \frac{|y_\ell|^2}{32\pi^2}
	\left[
		\left( 1 + \frac{k g_Y}{2g_X} \right) \cos \alpha_Z -
		\frac{g_2 \sqrt{1-k^2} \sin \alpha_Z}{g_X \cos \theta_W}
		\left( -\frac{1}{2} + \sin^2 \theta_W 	
		\right)
	\right]	g_B \, ,
\end{eqnarray}
\begin{eqnarray}		
B^Z_k (q^2)	&=&	
	\left[
	\frac{\sqrt{1-k^2}}{4\cos \theta_W} \cos \alpha_Z +
	\frac{k g_Y}{4g_2} \sin \alpha_Z
	\right]	
	- \, \frac{|y_\ell|^2 g_X \sin \alpha_Z}{32\pi^2 g_2} f_B
	\nonumber	\\
&&	
	+ \, \frac{|y_\ell|^2}{32\pi^2}
	\left[
	\frac{\sqrt{1-k^2} \cos \alpha_Z}{\cos \theta_W}
	\left( -\frac{1}{2} + \sin^2 \theta_W \right) +
	\left( 1+ \frac{k g_Y}{2 g_X} \right) 
	\frac{g_X \sin \alpha_Z}{g_2}
	\right]	g_B \, .
\end{eqnarray}
In these above formulas, the loop functions $f_A$, $g_A$, $f_B$, and $g_B$ are given by
\begin{eqnarray}
f_A	&=&
	\int dx dy dz \delta(1-x-y-z)	
	\left\{
	\frac{\ln \left[ (\tau z +x+y+ \delta x) (\tau z +x+y+ \delta y) \right] }{2}
	\right. 	\nonumber	\\
&&	 
	\qquad \qquad \qquad	
	\left.
	- \; \frac{m_\ell^2}{m_{\chi_r}^2}
	z(1-z)
	\left[ 
		\frac{1}{\tau z +x+y+ \delta x} + 
		\frac{1}{\tau z +x+y+ \delta y}
	\right]
	\right\}	\, ,	\\
%
g_A	&=&
	\int dx dy dz \delta(1-x-y-z)
	\left\{
	-\frac{ 
	\ln \left[ (\tau (x+y) +z)  (\tau(x+y) +z + \delta z) \right] }{2}
	\right.	\nonumber	\\
&&	
	\qquad \qquad
	\left.
	+ \, \frac{z^2 m_\ell^2 + xyq^2 + m_L^2}{2m_{\chi_r}^2}
	\left[
		\frac{1}{\tau (x+y) + z} +
		\frac{1}{\tau (x+y) + z + \delta z }
	\right]
	\right\}	\, ,	\\
%
f_B	&=&	
	\int dx dy dz \delta(1-x-y-z)
	\frac{\ln \left[ (\tau z +x+y+ \delta x) (\tau z +x+y+ \delta y) \right]}{2}	\, ,	\\
%
g_B	&=&	
	\int dx dy dz \delta(1-x-y-z) 
	\left\{
	\frac{\ln \left[ (\tau (x+y) +z) (\tau (x+y) +z+ \delta z) \right] }{2}
	\right.	\nonumber	\\
&&	
	\left.
	\qquad \qquad 
	+ \, \frac{z^2 m_\ell^2 - xyq^2 - m_L^2}{2 m_{\chi_r}^2}
	\left[
		\frac{1}{\tau (x+y) +z} +
		\frac{1}{\tau (x+y) +z+ \delta z}
	\right]
	\right\}	\, ,
\end{eqnarray}
as the results of the Feynman parameterization.
In these formulas, $\ell$ is one of the SM charge leptons $\{e, \mu, \tau\}$.

\section{Numerical analysis}

In the numerical analysis, we assume, for simplicity, that the new vectorlike leptons have sizable coupling with muons ($y_\mu$), while the corresponding coupling with electrons ($y_e$) is negligible.
Therefore, the set of relevant input parameters  are
\begin{eqnarray}
m_{\chi_r}, \,m_{Z'}, \,  
k, \, g_X, \, y_\mu , \,
\tau, \, \delta, \, 
A_{bs}, \, B_{bs} \, .
\end{eqnarray}
%


In this section, we consider the phenomenological constraints including the muon anomalous magnetic moments, and the rare semileptonic decays of $B$ mesons.
The current deviation between the experimental value and the SM prediction of muon $g-2$ is 
\cite{Tanabashi:2018oca,Abi:2021gix,Aoyama:2020ynm}
\begin{eqnarray}
\Delta a_\mu &\equiv &
	a_\mu^\text{exp} - a_\mu^\text{SM}
 \;=\;
 	(25.1 \pm 5.9)  
	\times 10^{-10} .
\label{amu}
\end{eqnarray}
The on-going E989 experiment will be able to reach a precision of 140 parts-per-billion \cite{Grange:2015fou}.
Assuming the same center value of $\Delta a_\mu$ as the current result (\ref{amu}), 
the projected difference between the SM prediction and the experimental value reads
\begin{eqnarray}
\Delta a_\mu^\text{projected} &=&
(25.1 \pm 4.6)
	\times 10^{-10} ,
\label{amu_proj}
\end{eqnarray}
corresponding to a 
5.5$\sigma$ deviation.

For the $B$-meson semileptonic decays, we take into account the branching ratios of the processes 
$B^+ \rightarrow K^+ \mu^+\mu^-$, 
$B^0 \rightarrow K^*(892)^{0} \mu^+\mu^-$, 
and the observables, $R_K$ and $R_{K^*}$, characterizing the violation of lepton flavor universality.
For the muon invariant mass in the region $q^2 = [1.1,\, 6.0]$  GeV$^2$, the 2$\sigma$ allowed ranges for the following $B$-meson observables are:
\begin{eqnarray}
1.050 \times 10^{-7} < BR(B^+ \rightarrow K^+ \mu^+\mu^-) < 1.322 \times 10^{-7}	\, , 
&&	
	\qquad	
	\cite{BR, Aaij:2012vr} 
\label{BR_B+}	
	\\
%
1.382 \times 10^{-7} < BR(B^0 \rightarrow K^{*0} \mu^+\mu^-) <  1.970 \times 10^{-7}	\, , 
&&	
	\qquad
	\cite{BR, Aaij:2016flj}	
\label{BR_B0}	
	\\
%
0.764 < R_K = \frac{BR(B^+ \rightarrow K^+ \mu^+\mu^-)}{BR(B^+ \rightarrow K^+ e^+e^-)} < 
0.934	\, ,
&&	
	\qquad
	\cite{RK_RKs, Aaij:2019wad}	
\label{RK}	
	\\
%
0.53 < R_{K^*} = \frac{BR(B^0 \rightarrow K^{*0} \mu^+\mu^-)}{BR(B^0 \rightarrow K^{*0} e^+e^-)} < 0.95	\, ,	
&&	
	\qquad
	\cite{RK_RKs, Aaij:2017vbb}
\label{RK*}	
\end{eqnarray}
according to the updated results from the LHCb experiment.
The branching ratios of the $B$-meson decay processes are calculated using the methods described in Refs. 
\cite{Altmannshofer:2012az, Wang:2017mrd, Bobeth:2008ij}.
The relevant form factors from Ref. \cite{Bharucha:2010im} are employed 
in our calculation.

The searches for the $Z'$ boson have been carried out at the LHC run II in various channels  where the resonance decays into 
dilepton \cite{ATLAS:2017eiz, ATLAS:2019erb, CMS:2021ctt},
diquark \cite{ATLAS:2018tfk, ATLAS:2020lks, CMS:2019gwf, CMS:2019xai, CMS:2019emo}, 
and $Zh$ \cite{ATLAS:2020pgp}
for a wide range of $m_{Z'}$ up to 6 TeV.
These analyses set the upper limits on the production cross section times branching ratios of the $Z'$ boson that in turn impose constraints on the
$Z'$ mass and its couplings to the SM particles.
In the BDW model, to address the muon $g-2$ anomaly, the loop-induced effective coupling between $Z'$ and muons needs to be large enough.
As a consequence, the $Z'$ boson decays dominantly to $\mu\bar{\mu}$ and $\nu_\mu \bar{\nu}_\mu$ \cite{Belanger:2015nma}.
Therefore, among these constraints, those from the dimuon searches is the most severe for the BDW model.

The constraints on the kinetic mixing coefficient $k$ and $m_{Z'}$ from electroweak precision tests and other various experimental data from channels like
the $h \rightarrow Z Z'$ and 
$h \rightarrow Z' Z'$ decays,
the Drell-Yan $Z'$ production were studied in Ref. 
\cite{Curtin:2014cca}.
According to that, for a wide range of the $Z'$-boson mass below $\mathcal{O}(1)$ TeV, the current limit for the gauge kinetic mixing coefficient is 
$k \lesssim \mathcal{O}(10^{-2})$.
For small mass region of the $U(1)_X$ gauge boson below 10 GeV, the analyses by the BarBar Collaboration \cite{Lees:2014xha}
and the KLOE-2 Collaboration \cite{Anastasi:2018azp}
indicate the most stringent upper bound on the kinetic mixing coefficient.
Recently, the CMS Collaboration investigated the muon pair production channel at the LHC with the center of mass energy $\sqrt{s} = 13$ TeV in the search for a narrow resonance 
\cite{Sirunyan:2019wqq}.
Similar analysis was also studied by the LHCb Collaboration \cite{Aaij:2019bvg}.
These results show a severe constraint on the kinetic mixing coefficient with a $Z'$ boson lighter than 200 GeV.
The approximated upper bound on $k$ for various range of $m_{Z'}$ is summarized as follows
\begin{eqnarray}
k & \lesssim & 
\begin{cases}
	\,
	10^{-3},
&	\text{for} \quad
	m_{Z'} < 10 \text{ GeV},
	\hspace{2.8cm}	
	\cite{Lees:2014xha, Anastasi:2018azp}
	\\
	\,	
	1.5 \times 10^{-3},
&	\text{for} \quad
	10.6 \text{ GeV} < m_{Z'} < 30 \text{ GeV},	
	\hspace{1cm}
	\cite{Aaij:2019bvg}
	\\
	\,
	2\times 10^{-3},
&	\text{for} \quad
	30 \text{ GeV} < m_{Z'} < 75 \text{ GeV},
	\hspace{1.35cm}	
	\cite{Sirunyan:2019wqq}
	\\
	\,
	4\times 10^{-3},
&	\text{for} \quad
	110 \text{ GeV} < m_{Z'} < 200 \text{ GeV}	.
	\hspace{1cm}
	\cite{Sirunyan:2019wqq}
\end{cases}
\label{upper_bound_k}
\end{eqnarray}


At the LEP experiment, the search for sleptons used the channels with the same final states as those coming from the vectorlike leptons.
Therefore, they
provide a lower limit for the mass of charged vectorlike leptons
\cite{Abdallah:2003xe}:
\begin{eqnarray}
m_L	& \gtrsim & 
	100 \text{ GeV}.
\label{LEP_constraint}
\end{eqnarray}
Similarly, the LHC constraint on the vectorlike lepton masses can be derived from the data of the slepton searches at the ATLAS and CMS experiments at 13 TeV
\cite{Aad:2014vma}.
According to that, the vectorlike leptons must satisfy either $m_L \gtrsim \mathcal{O}(1)$ TeV, or
\begin{eqnarray}
m_L - m_{\chi_r} \lesssim 60 \text{ GeV}.
\label{small_mass_splitting}
\end{eqnarray} 
To explain the muon $g-2$ while keeping the coupling $y_\mu$ in the perturbative regime, the vectorlike leptons must be light enough.
Therefore, the scenario with a small gap between $m_L$ and $m_{\chi_r}$ is preferable.
For $m_{\chi_r} \sim \mathcal{O}(100)$ GeV, the condition (\ref{small_mass_splitting}) implies
$\tau \sim \mathcal{O}(1)$.
Assuming that the particle $\chi_r$ is stable, the condition (\ref{small_mass_splitting})
becomes
\begin{eqnarray}
0 < m_{L} - m_{\chi_r} \lesssim 60 \text{ (GeV)}.
\label{small_mass_splitting0}
\end{eqnarray}
The parameter region with such compressed mass spectra is subjected to a constraint from the recent analysis by the ATLAS Collaboration \cite{ATLAS:2019lng}.

\begin{figure}[h]
\begin{center}
\includegraphics[scale=0.5]{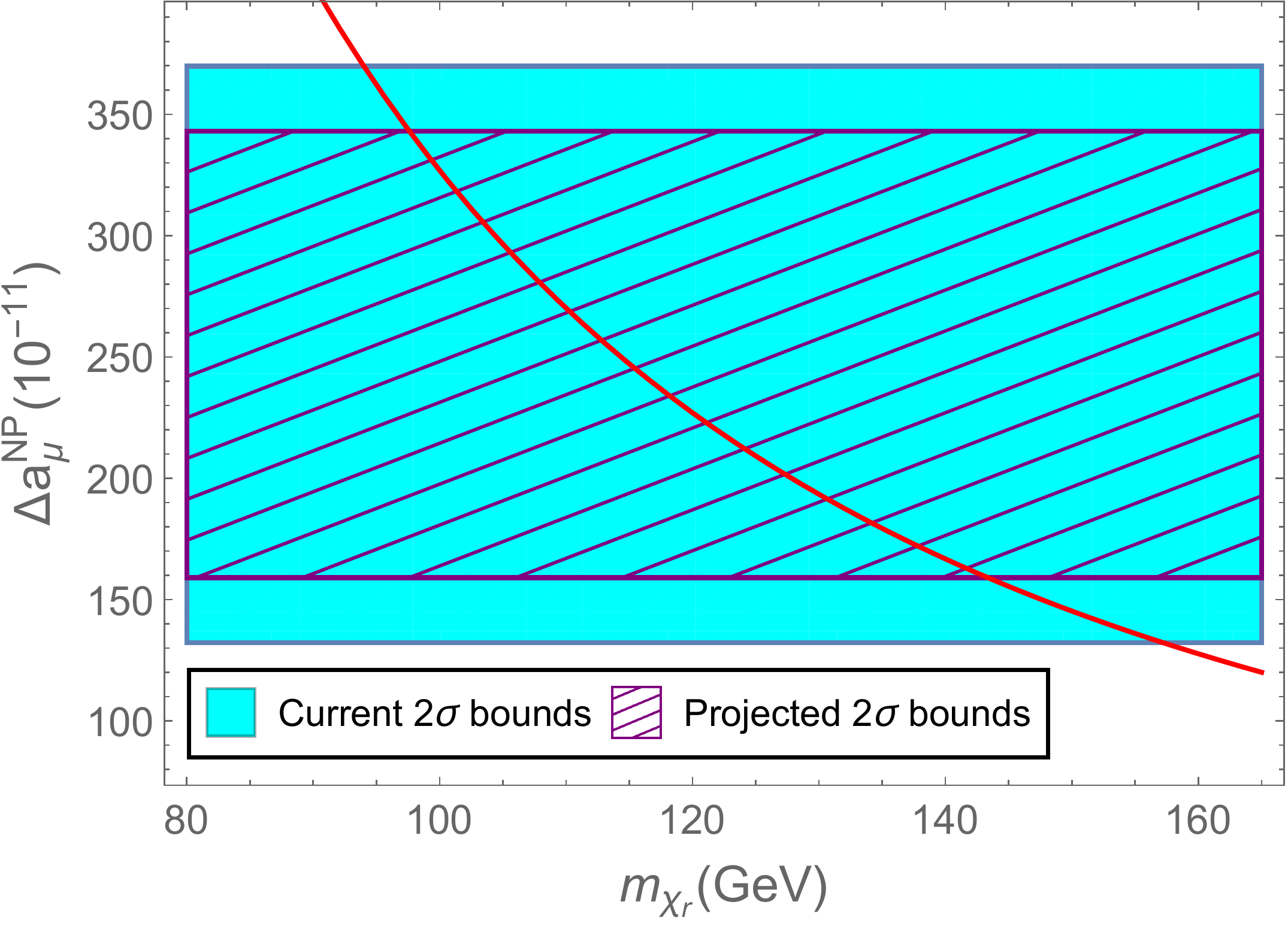}
\caption{The new physics contributions to the muon anomalous magnetic moment as a function of $m_{\chi_r}$ for the case of 
$\tau=1.78$, $\delta=1$, $y_\mu = 3$,
and $k=0$.}
\label{amu_mchi}
\end{center}
\end{figure}

\textbf{Constraints on $m_{\chi_r}$: }

Since the mass of the scalar field $\chi_r$ only appears as an independent parameter in Eq. (\ref{amu_k}), 
the bounds (\ref{amu}) induce a constraint on the parameter $m_{\chi_r}$ to explain the observed muon $g-2$.
In Figure \ref{amu_mchi}, we show the dependence of $\Delta a_\mu^\text{NP}$ on $m_{\chi_r}$ for fixed values of other inputs, 
$\tau=1.78$, $\delta=1$, $y_\mu = 3$,
and $k=0$.
In this case, the current bounds on the muon $g-2$ yield the $2\sigma$ allowed range for the $\chi_r$ mass to be 
$94 \text{ GeV} \lesssim m_{\chi_r} \lesssim 157 \text{ GeV}$.%
\footnote{
Note that when $m_{\chi_r}$ is larger than this upper limit, although the contribution of new physics to the muon $g-2$ is not large enough to explain the measured value at the level of $2\sigma$, the model is not ruled out completely since its predictions are still in agreement with the SM ones.
}
In the near future, when the measurement at the E989 experiment is completed, we expect that this range will be improved.
The projected bounds in Eq. (\ref{amu_proj}) imply more severe $2\sigma$ limits for this parameter that are 
$98 \text{ GeV} \lesssim m_{\chi_r} \lesssim 143 \text{ GeV}$.
In the subsequent analysis regarding the muon $g-2$, we choose 
$m_{\chi_r} = 120$
GeV as a representative value.

\textbf{Constraints on 
$(A_{bs},B_{bs})$ plane:}

\begin{figure}[h]
\begin{center}
\includegraphics[scale=0.65]{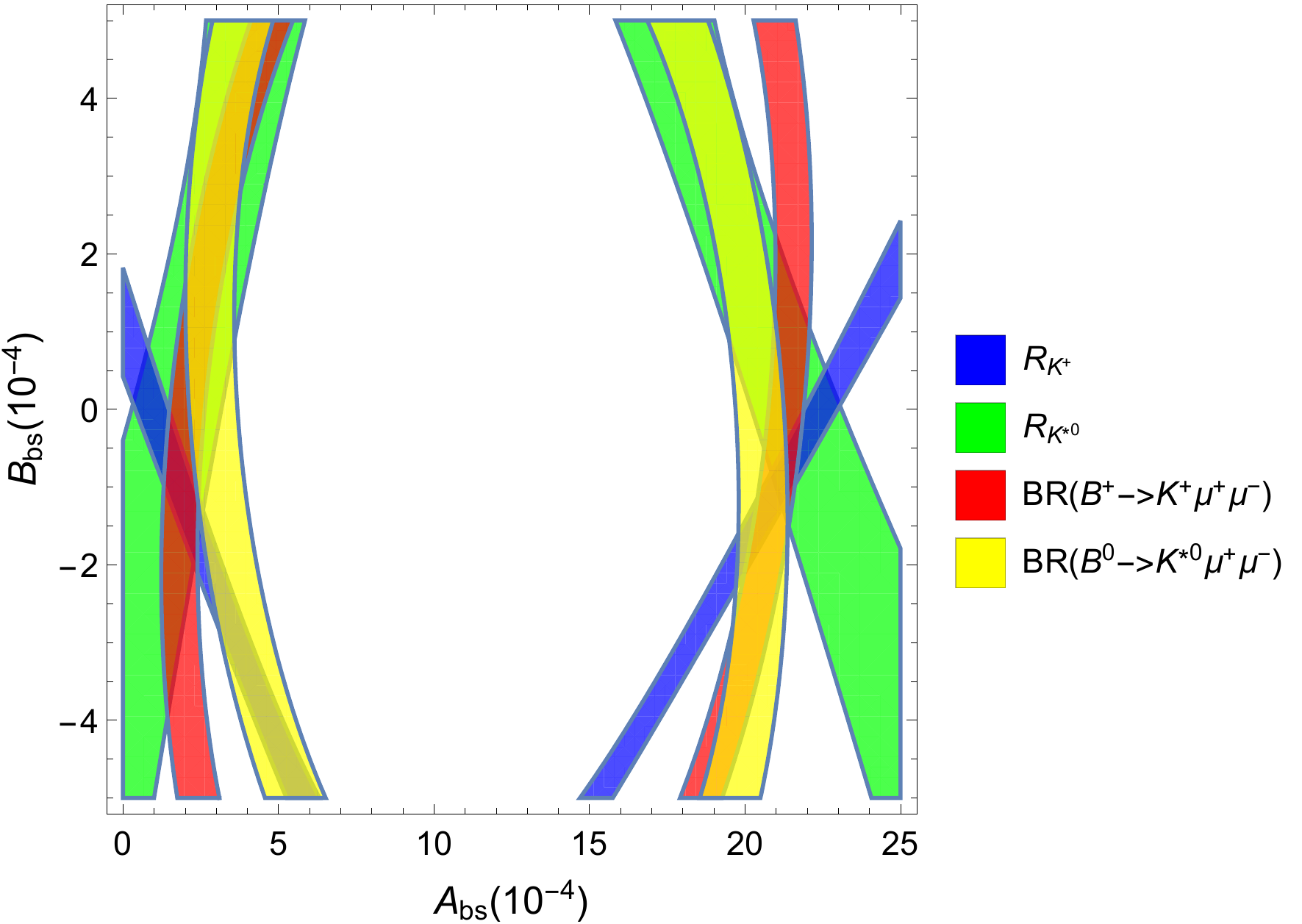}
\caption{
Phenomenological constraints on the $(A_{bs},B_{bs})$ plane
for the case of 
$m_{Z'}=300$ GeV, 
$y_\mu = 3$,  
$g_X = 3$, 
$\tau=1.78$, 
$\delta=1$, 
and $k=0$.
}
\label{AB_k0}
\end{center}
\end{figure}

\begin{figure}[h!]
\begin{minipage}{0.48\textwidth}
\begin{flushleft}
\includegraphics[scale=0.42]{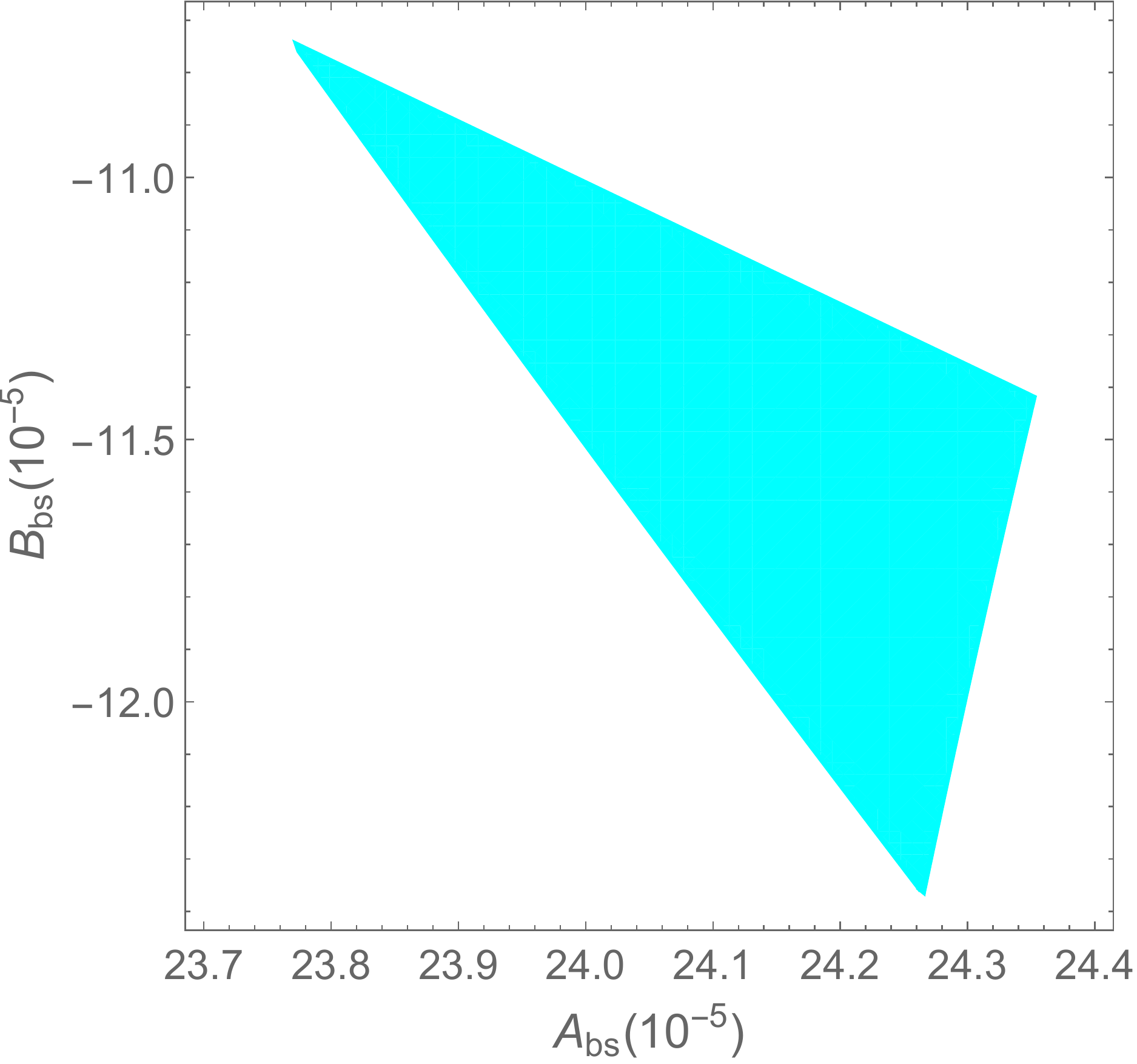}
\caption{
The left parameter region on the $(A_{bs},B_{bs})$ plane
for the case of 
$m_{Z'}=300$ GeV, 
$y_\mu = 3$,  
$g_X = 3$, 
$\tau=1.78$, 
$\delta=1$, 
and $k=0$.
}
\label{AB_k0_allowed_left}
\end{flushleft}
\end{minipage}
\hspace{0.5cm}
%
\begin{minipage}{0.48\textwidth}
\begin{flushright}
\includegraphics[scale=0.4]{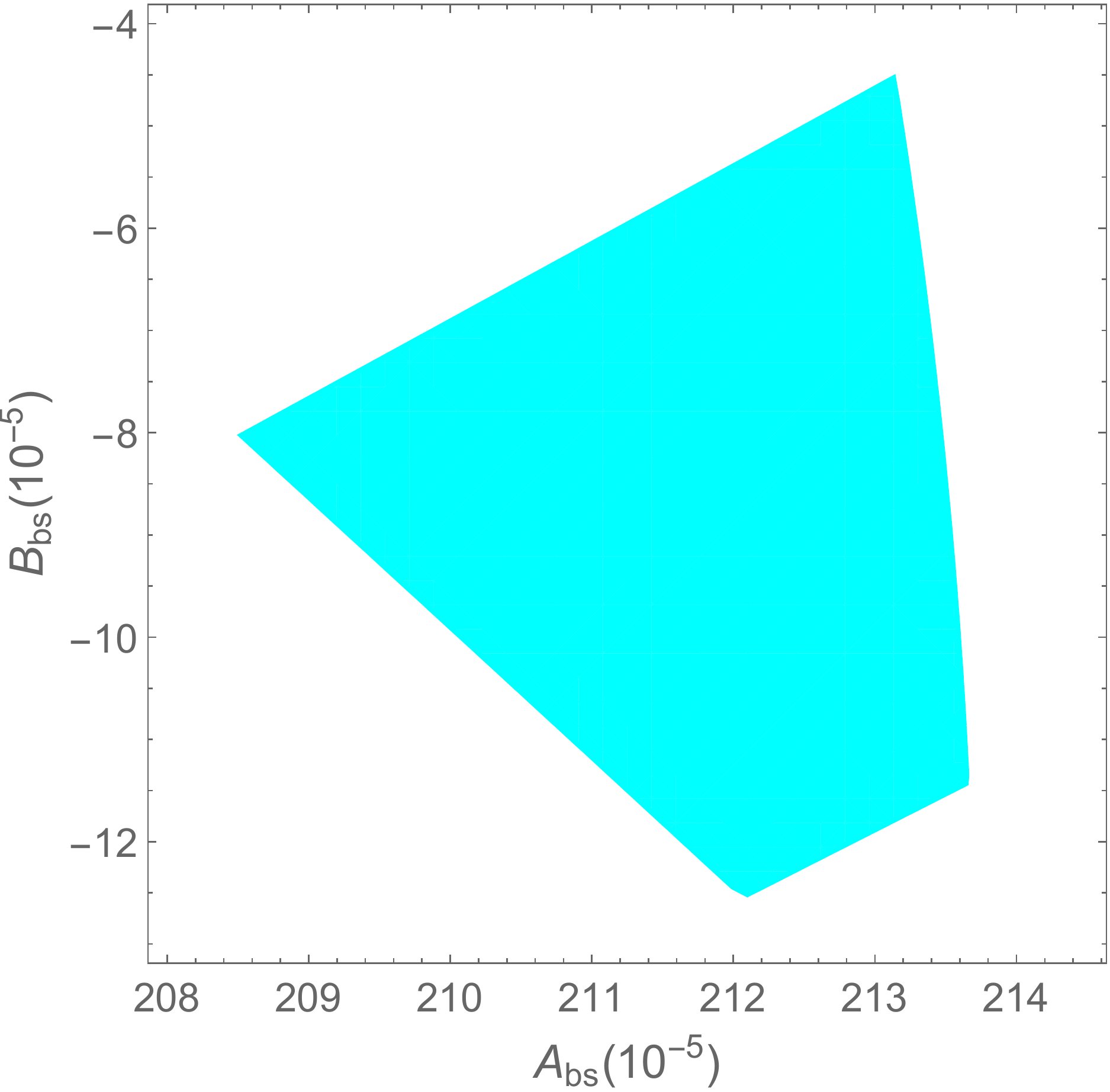}
\caption{
The right parameter region on the $(A_{bs},B_{bs})$ plane
for the case of 
$m_{Z'}=300$ GeV, 
$y_\mu = 3$,  
$g_X = 3$, 
$\tau=1.78$, 
$\delta=1$, 
and $k=0$.
}
\label{AB_k0_allowed_right}
\end{flushright}
\end{minipage}
\end{figure}

\begin{figure}[h!]
\begin{minipage}{0.48\textwidth}
\begin{flushleft}
\includegraphics[scale=0.42]{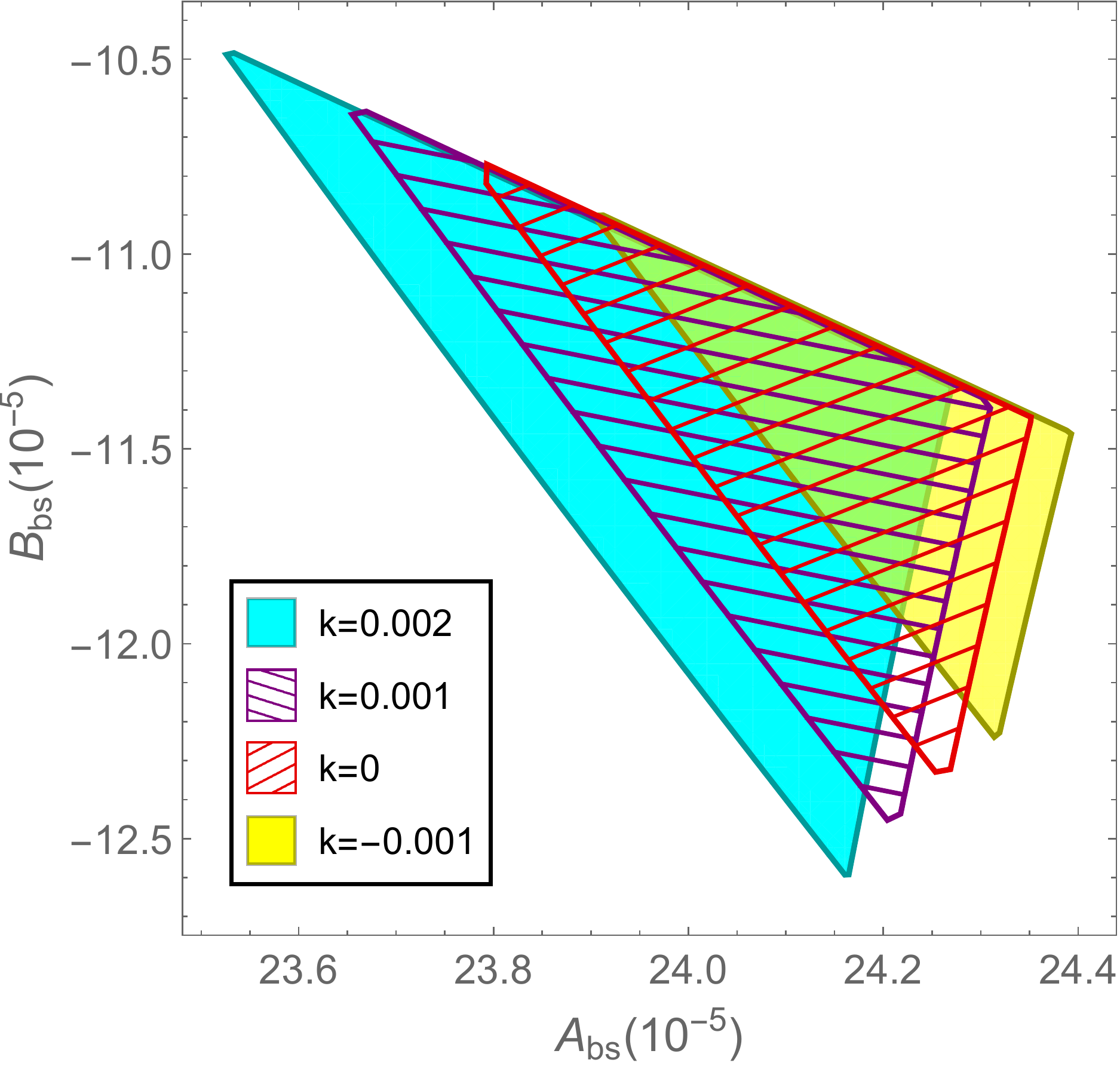}
\caption{
The left parameter region on the $(A_{bs},B_{bs})$ plane
for the case of 
$m_{Z'}=300$ GeV, 
$y_\mu = 3$,  
$g_X = 3$, 
$\tau=1.78$, 
$\delta=1$, 
and 
$k=-0.001$, 0, 
$0.001$, 
$0.002$.   
}
\label{AB_k_overlap_L}
\end{flushleft}
\end{minipage}
\hspace{0.5cm}
%
\begin{minipage}{0.48\textwidth}
\begin{flushright}
\includegraphics[scale=0.42]{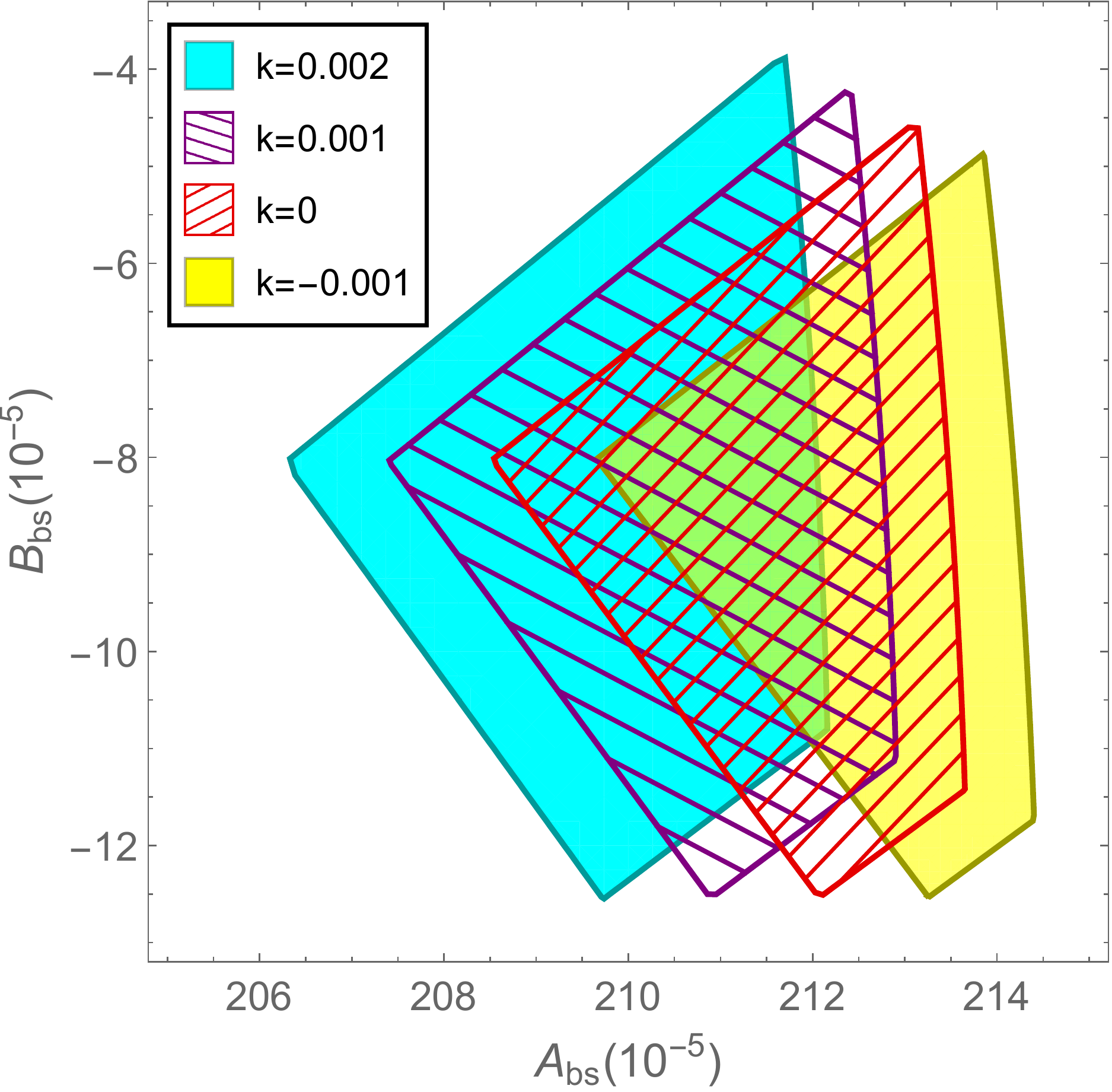}
\caption{
The right parameter region on the $(A_{bs},B_{bs})$ plane
for the case of 
$m_{Z'}=300$ GeV, 
$y_\mu = 3$,  
$g_X = 3$, 
$\tau=1.78$, 
$\delta=1$, 
and 
$k=-0.001$, 0, 
$0.001$, 
$0.002$.   
}
\label{AB_k_overlap_R}
\end{flushright}
\end{minipage}
\end{figure}

For the parameters $A_{bs}$ and $B_{bs}$, they are constrained by the measurements of the branching fractions 
of the semileptonic decays 
$B^+ \rightarrow K^+ \mu^+ \mu^-$, and
$B^0 \rightarrow K^{*0} \mu^+ \mu^-$ 
(Eqs. (\ref{BR_B+})-(\ref{BR_B0})), 
as well as 
the ratios measuring the lepton universality violation 
$R_K$ and $R_{K^*}$ 
(Eqs.(\ref{RK})-(\ref{RK*})).
In Figure \ref{AB_k0}, these constraints (represented by the red, yellow, blue and green colors, respectively) at the level of $2\sigma$ are depicted on the $(A_{bs}, B_{bs})$ plane for fixed values of other parameters
$m_{Z'}=300$ GeV, 
$y_\mu = 3$, 
$g_X = 3$, 
$\tau=1.78$, and 
$\delta=1$, 
in the case of vanishing kinetic mixing.
The strips corresponding to each of these constraints appear approximately in the elliptical forms. It is due to the fact that the relevant observables are quadratic functions of the Wilson coefficients, $C_9^{(')\text{NP}}$ and $C_{10}^{(')\text{NP}}$, that in turns are proportional to the first order of the parameters $A_{bs}$ and $B_{bs}$.
%
%
%
%
Here, we see that there are two overlap regions on the left and on the right of the figure that satisfy all these constraints.
%
They are extracted and shown separately in Figures \ref{AB_k0_allowed_left} and \ref{AB_k0_allowed_right}.
We observe that the allowed ranges for $A_{bs}$ and $B_{bs}$ in 
the left viable region are
$23.77 \times 10^{-5} \lesssim A_{bs} \lesssim 24.36 \times 10^{-5}$, and
$-12.38 \times 10^{-5} \lesssim B_{bs} \lesssim -10.74 \times 10^{-5}$, respectively.
For the right viable region, the corresponding limits are
$208.5 \times 10^{-5} \lesssim A_{bs} \lesssim 213.7 \times 10^{-5}$, and
$-12.55 \times 10^{-5} \lesssim B_{bs} \lesssim -4.46 \times 10^{-5}$.
From the magnitudes of these two parameters, it is clearly that all the relevant FCNC processes are very much suppressed.

In Figure \ref{AB_k_overlap_L}, 
the left regions satisfying all the considered constraints from $B$-meson decays are depicted for different scenarios with the kinetic mixing coefficient to be
$k=-0.001$, 0, $0.001$, and $0.002$.
They are shown as the yellow, red slashed, purple back-slashed, and cyan regions in the plot.
We observe that the viable region 
shifts to the left toward smaller values of $A_{bs}$, and
increases its area 
when the kinetic mixing coefficient becomes larger.
As the consequence, the windows for the parameters $A_{bs}$ and $B_{bs}$ become more relaxed for larger values of $k$.
In Figure 
\ref{AB_k_overlap_R}, the situation for the right regions is shown when changing the value of the kinetic mixing coefficient, namely
$k = -0.001, 0, 0.001$, and 0.002.
We observe that the lower bound of $B_{bs}$ in the right region ($\sim 12.55 \times 10^{-5}$) remains almost unchanged for various values of $k$, while its upper bound slightly increases for larger $k$.
Regarding to the parameter $A_{bs}$, a larger value of $k$ shifts its allowed range toward the left.
Meanwhile, the width of this range becomes slightly larger when increasing $k$.
Since $A_{bs}$ in the right region is about $\mathcal{O}(10)$ times larger than that in the left region, it enhances the cross section 
$\sigma(pp \rightarrow Z'
 \rightarrow \mu \bar{\mu})$
via the $b\bar{s}/s\bar{b}$ annihilations roughly by a factor of $\mathcal{O}(10^2)$.
Taking into account the constraint in Ref. \cite{ATLAS:2019erb}, we find that the right region is excluded while the left one is allowed.


\textbf{Constraints on 
$(\tau,\delta)$ plane:}

\begin{figure}[h!]
\begin{minipage}{0.48\textwidth}
\begin{flushleft}
\includegraphics[scale=0.56]{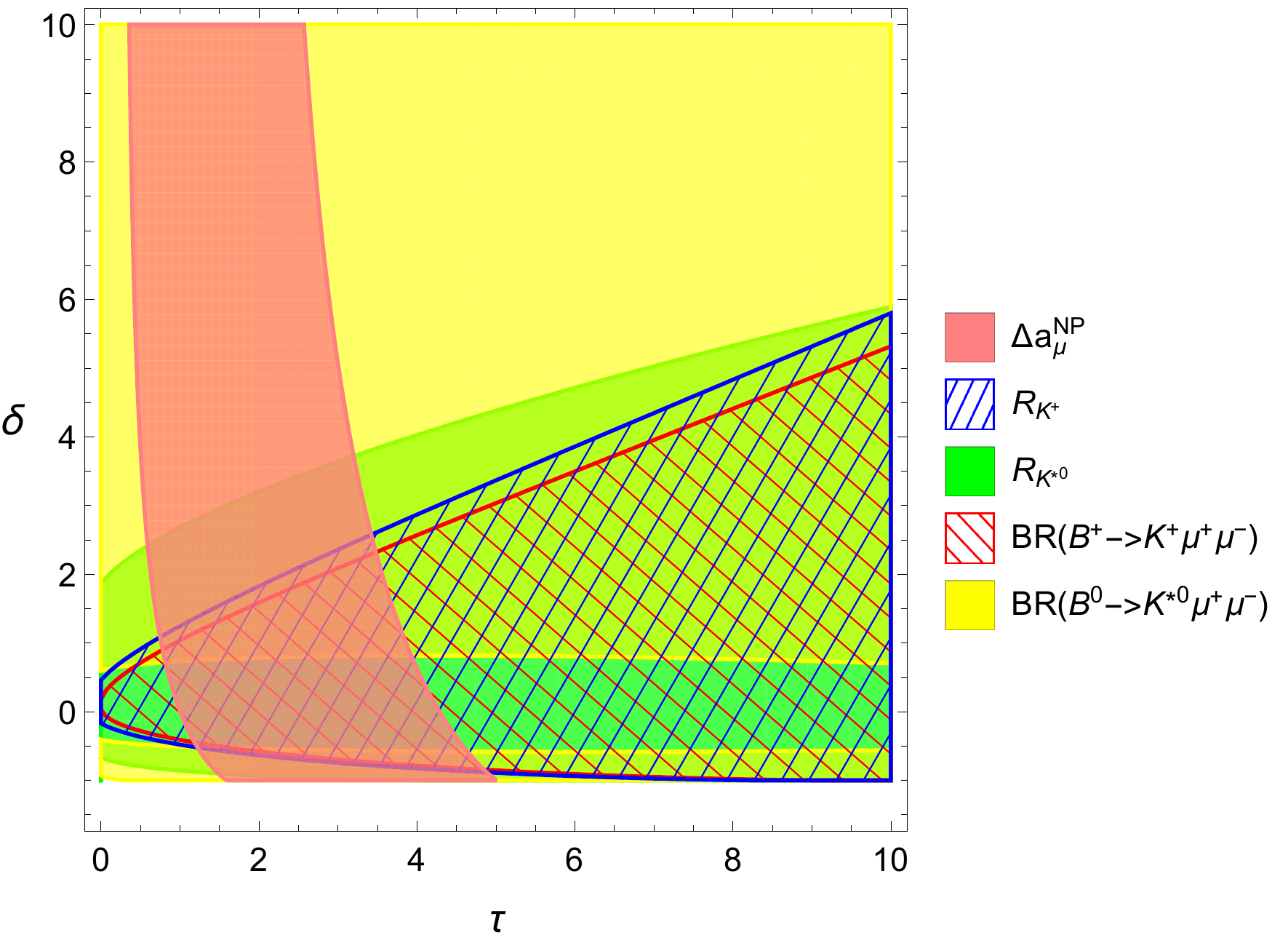}
\caption{
Phenomenological constraints on the $(\tau,\delta)$ plane
for the case of 
$m_{\chi_r}=120$ GeV, 
$m_{Z'}=300$ GeV, 
$y_\mu = 3$, 
$g_X = 3$, 
$A_{bs}=24.2 \times 10^{-5}$, 
$B_{bs}=-11.5 \times 10^{-5}$, 
and $k=0$.}
\label{td_k0}
\end{flushleft}
\end{minipage}
\hspace{0.5cm}
%
\begin{minipage}{0.48\textwidth}
\begin{flushright}
\includegraphics[scale=0.38]{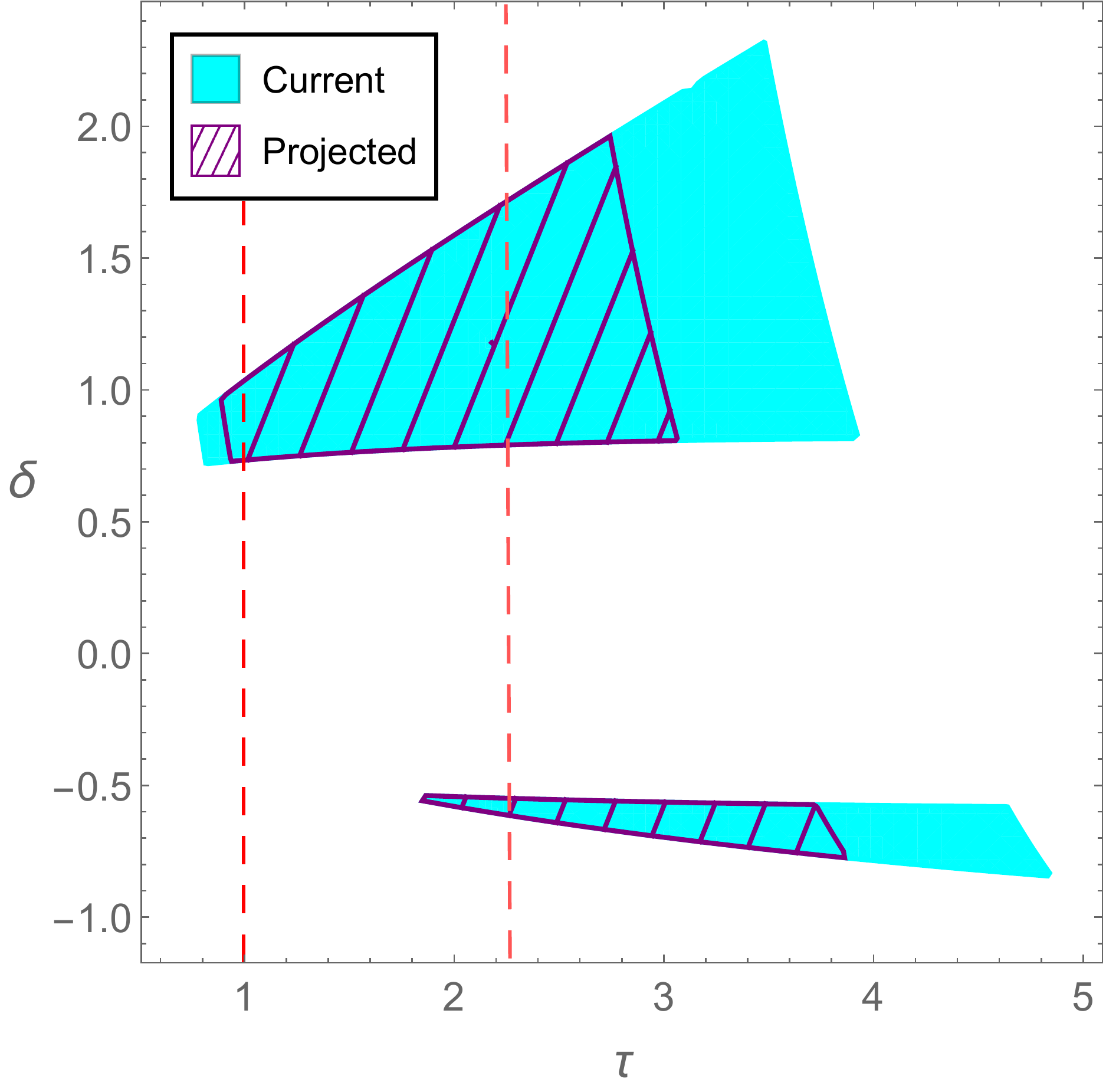}
\caption{
Viable parameter region on the $(\tau,\delta)$ plane
for the case of 
$m_{\chi_r}=120$ GeV, 
$m_{Z'}=300$ GeV, 
$y_\mu = 3$, 
$g_X = 3$, 
$A_{bs}=24.2 \times 10^{-5}$, 
$B_{bs}=-11.5 \times 10^{-5}$, 
and $k=0$.
The hatched region corresponds to the projected result after the E989 experiment.
The region between the two vertical red dashed lines satisfies the constraint in Eq. (\ref{small_mass_splitting0}).
}
\label{td_proj}
\end{flushright}
\end{minipage}
\end{figure}

The parameters $\tau$ and $\delta$ involve in all the considered observables.
Since $\delta$ is defined by the Eq. (\ref{delta_define}), it must satisfy the theoretical condition $\delta \geqslant -1$.
In Figure \ref{td_k0}, we show how the constraints on 
$\Delta a_\tau^\text{NP}$, $R_{K}$, 
$R_{K^*}$, 
$BR(B^+ \rightarrow K^+ \mu^+ \mu^-)$, and
$BR(B^0 \rightarrow K^{*0} \mu^+ \mu^-)$ at the level of 2$\sigma$ affect the 
$(\tau,\delta)$ plane in the case with a vanishing kinetic mixing coefficient and fixed values of other inputs:
$m_{\chi_r}=120$ GeV, 
$m_{Z'} = 300$ GeV, 
$y_\mu = 3$, 
$g_X = 3$,
$A_{bs}=24.2 \times 10^{-5}$, 
and 
$B_{bs}=-11.5 \times 10^{-5}$. 
We observe that most of the interested parameter region satisfies the constraint on 
$BR(B^0 \rightarrow K^{*0} \mu^+ \mu^-)$.
The region with $-0.5 \lesssim \delta \lesssim 0.5$ results in a more frequent decay of $B^0$ mesons into $K^{*0}$ and a pair of muons that is above the upper limit.
Therefore, this region is excluded.
The bounds on new physics contributions to the muon $g-2$ require $0.3 \lesssim \tau \lesssim 3.2$
when $\delta \gtrsim 3.7$.
It is because, for such large values of $\delta$, the second term in the squared bracket in Eq. (\ref{amu_k}) is negligible, and $\Delta a_\mu^\text{NP}$ depends mostly on the first term.
As a consequence, the bounds on $\Delta a_\mu^\text{NP}$ specify the allowed range for $\tau$ that is almost independent of 
$\delta$.
However, this region with large $\delta$ is roughly excluded by the experimental data on 
$R_{K^{*0}}$.
For $\delta$ smaller than about 3.7, 
the dependence of the allowed range of $\tau$ on $\delta$ becomes clearer. 
From this figure, beside the constraint on the muon $g-2$, we see that those on 
the semileptonic branching ratios of the $B^0$ 
and $B^+$
 mesons play an important role in determining the 2$\sigma$ allowed parameter region.
In Figure \ref{td_proj}, the regions satisfying all these current bounds is shown in the cyan color.
When taking into account the projected result from the E989 experiment, the viable parameter regions significantly reduce.
These regions are depicted by the hatched areas in Figure \ref{td_proj}.
For $m_{\chi_r} = 120$ GeV,
the LEP constraint in Eq.  (\ref{LEP_constraint}) is automatically satisfied when we assume the LHC constraint in Eq. (\ref{small_mass_splitting0}). 
The latter leads to the constraint on $\tau$ that is shown as the parameter region
 between the two vertical red dashed lines in Figure \ref{td_proj}.
These bounds for $\tau$ from Eq. (\ref{small_mass_splitting0}) are actually more severe than those expected at the E989 experiment in the near future.
We see that there are two separated viable regions corresponding to positive and negative values of $\delta$.
For the chosen set of other inputs, the allowed ranges for the two parameters
in the former region are 
$1.00 < \tau \lesssim 2.25$ and 
$0.73 \lesssim \delta \lesssim 1.72$, 
while those ranges in the latter one are
$1.82 \lesssim \tau \lesssim 2.25$ and 
$-0.62 \lesssim \delta \lesssim -0.53$.
The search for the electroweak production of supersymmetric particles with compressed mass spectra at the LHC 13 TeV \cite{ATLAS:2019lng} excludes the following range for the $\tau$ parameter given that $m_{\chi_r} = 120$ GeV:
\begin{eqnarray}
1.02 \; \lesssim \; 
	\tau 
	\; \lesssim \; 1.43 .
\label{excluded}
\end{eqnarray}
Therefore, while the above negative-$\delta$ region remains intact, Eq. (\ref{excluded}) leads to two distinct positive-$\delta$ regions for the $\tau$ parameter:
(\textit{i}) $1.00 < \tau \lesssim 1.02$, and
(\textit{ii}) $1.43 \lesssim \tau \lesssim 2.25$.

\begin{figure}[h]
\center
\includegraphics[scale=0.4]{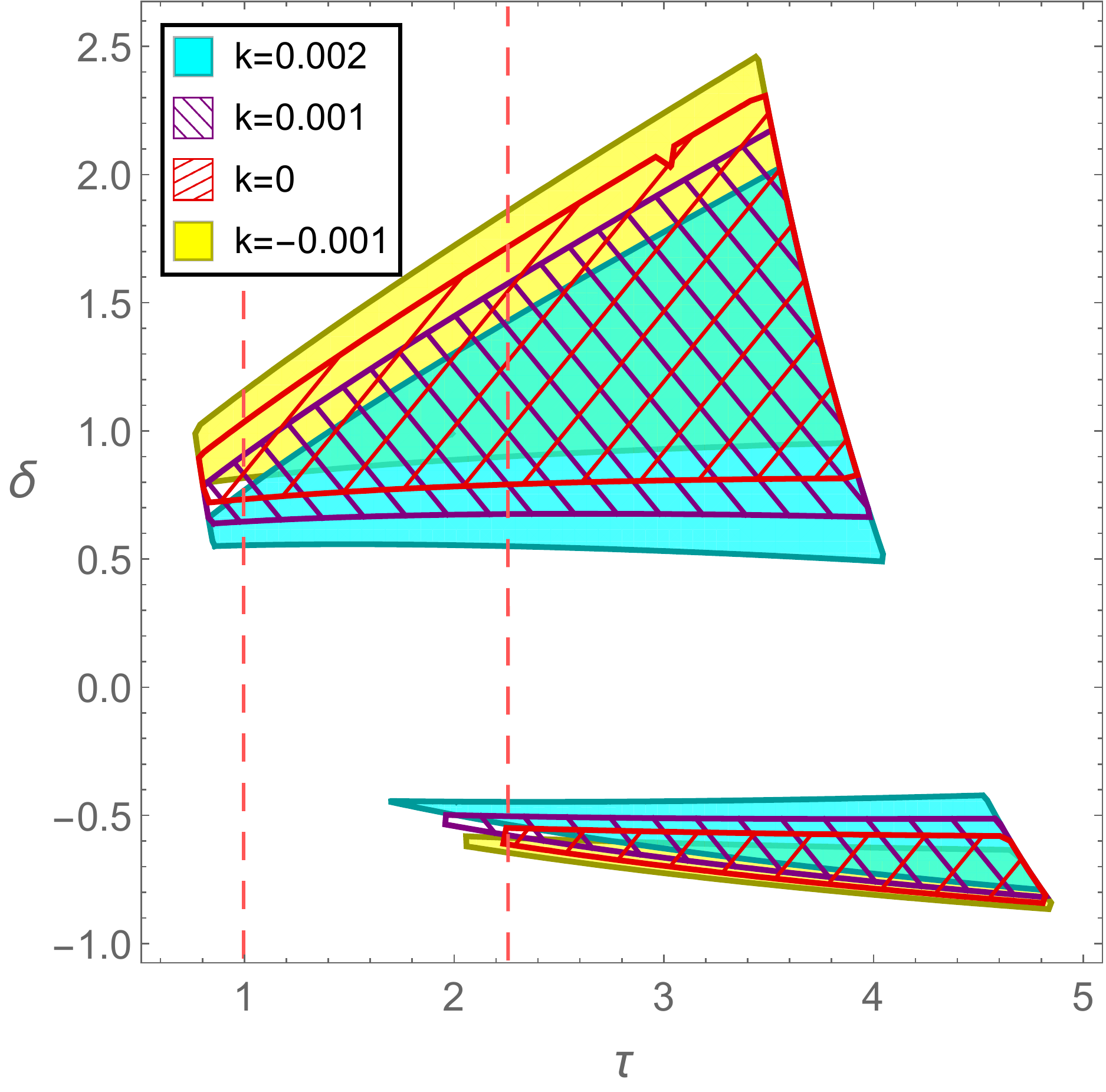}
\caption{
Viable parameter regions on the $(\tau,\delta)$ plane
for the case of 
$m_{\chi_r}=120$ GeV, 
$m_{Z'}=300$ GeV, 
$y_\mu = 3$, 
$g_X = 3$, 
$A_{bs}=24.2 \times 10^{-5}$, 
$B_{bs}=-11.5 \times 10^{-5}$, 
and various values of the kinetic mixing coefficient 
$k=-0.001, 0, 0.001, 0.002$.
The region between the two vertical red dashed lines satisfies the constraint in Eq. (\ref{small_mass_splitting0}).
}
\label{td_k_overlap}
\end{figure}

For nonzero kinetic mixing coefficients, the allowed parameter space gradually changes.
In Figure \ref{td_k_overlap}, the allowed regions on the $(\tau, \delta)$ plane are plotted for various values of the kinetic mixing coefficient, namely $k=-$%
0.001, 0, 0.001, and 0.002. 
Similar to Figure \ref{td_proj}, the region between the two vertical red dashed lines satisfies the constraint in Eq. (\ref{small_mass_splitting0}).
Here, we see that when increasing $k$, the viable range for $|\delta|$ moves toward smaller values, leading to a narrower gap between the positive-$\delta$ and the negative-$\delta$ regions.
It is noteworthy that the area of the positive-$\delta$ region is much larger than the negative-$\delta$ one implying that the scenario with lighter $\chi_r$ is more favorable than the one with lighter $\chi_i$.
Relevant to the parameter $\tau$, when the value of $k$ increases, the range of $\tau$ in the positive-$\delta$ region
remains unchanged due to the constraint (\ref{small_mass_splitting0}),
while the range of $\tau$ in the negative-$\delta$ region expands to the left.


\textbf{Constraints on the 
$(g_X, y_\mu)$ plane:}

\begin{figure}[h]
\begin{minipage}{0.48\textwidth}
\begin{flushleft}
\includegraphics[scale=0.57]{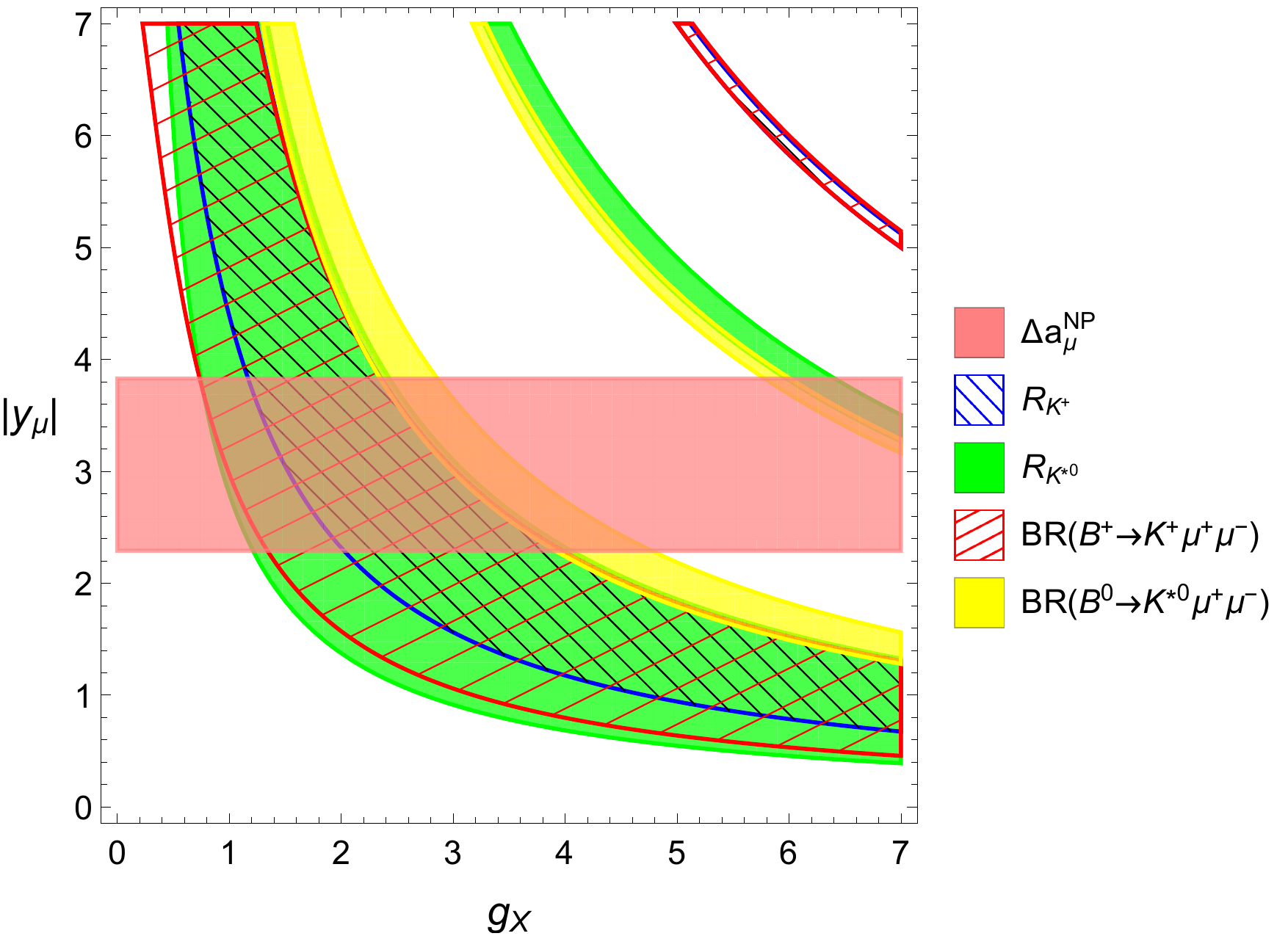}
\caption{
Phenomenological constraints on the $(g_X,y_\mu)$ plane
for the case of 
$m_{\chi_r}=120$ GeV, 
$m_{Z'}=300$ GeV, 
$A_{bs}=24.2 \times 10^{-5}$, 
$B_{bs}=-11.5 \times 10^{-5}$, 
$\tau=1.78$, 
$\delta=1$, 
and $k=0$.
}
\label{gy_k0}
\end{flushleft}
\end{minipage}
\hspace{0.5cm}
%
\begin{minipage}{0.48\textwidth}
\begin{flushright}
\includegraphics[scale=0.3]{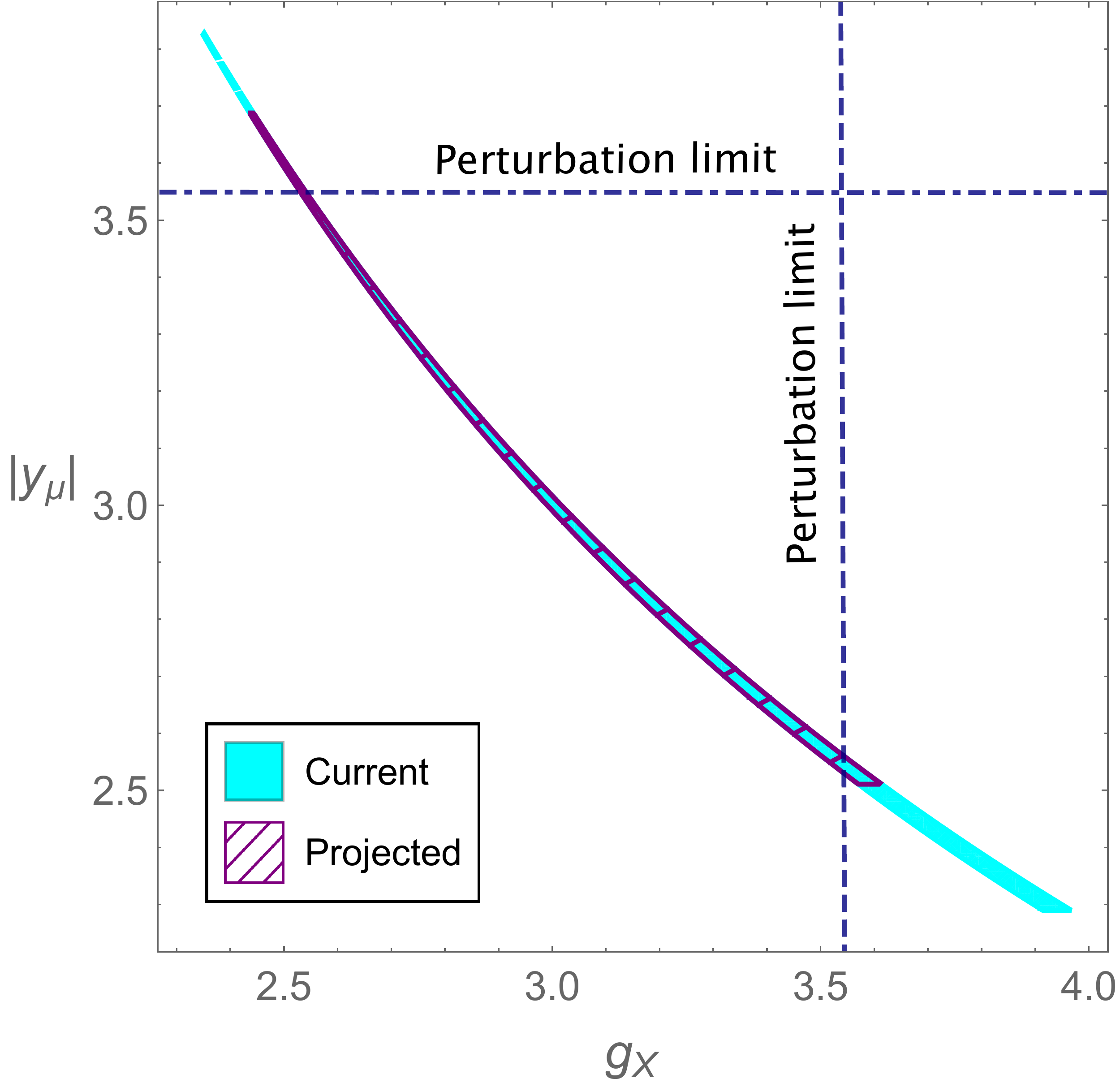}
\caption{
Viable parameter region on the $(g_X,y_\mu)$ plane
for the case of 
$m_{\chi_r}=120$ GeV, 
$m_{Z'}=300$ GeV, 
$A_{bs}=24.2 \times 10^{-5}$, 
$B_{bs}=-11.5 \times 10^{-5}$, 
$\tau=1.78$, 
$\delta=1$, 
and $k=0$.
The hatched region corresponds to the projected result after the E989 experiment.
}
\label{gy_k0_proj}
\end{flushright}
\end{minipage}
\end{figure}

In Figure \ref{gy_k0}, the phenomenological constraints are plotted on the $(g_X, y_\mu)$ plane in the case of no kinetic mixing and fixed values of other inputs, namely
$m_{\chi_r}=120$ GeV, 
$m_{Z'}=300$ GeV, 
$A_{bs}=24.2 \times 10^{-5}$, 
$B_{bs}=-11.5 \times 10^{-5}$, 
$\tau=1.78$, 
and
$\delta=1$. 
Since the muon $g-2$ does not depend on the coupling $g_X$, the corresponding allowed region has the form of a horizontal band with 
$2.27 \lesssim |y_\mu| \lesssim 
3.84$.
The parameter regions satisfying the constraints for other observables ($R_K$, $R_{K^*}$, and the branching ratios of $B^+$ and $B^0$ decays)
 have the hyperbolic forms.
This is due to the Wilson coefficients $C_{9,10}^{(')\text{NP}}$ that define the correlation between $g_X$ and $y_\mu$ for given values of the branching ratios.
The viable region determined by all of the five phenomenological constraints is shown separately as the cyan region in Figure \ref{gy_k0_proj}.
We see that the overlapping region is a thin strip in the hyperbolic form.
Here, the viable range for the parameter $g_X$ is
$2.35 \lesssim g_X \lesssim 3.97$.
In the near future, the E989 experiment will impose a more severe constraint on the 
$(g_X,y_\mu)$ plane which is shown in the hatched region.
To explain the anomalies on the $B$-meson decays, the new Yukawa coupling, $y_\mu$, and the $U(1)_X$ gauge coupling, $g_X$, are required to have large values of $\mathcal{O}(1)$.
Taking into account the perturbation limits for 
these two parameters shown as the horizontal dot-dashed and the vertical dashed lines
in this figure, we observe that this theoretical condition excludes a large portion of the allowed parameter region.
Moreover, the perturbation condition on $g_X$ indirectly determines the lower bound for $y_\mu$, and the perturbation condition on $y_\mu$ indirectly determines the lower bound for $g_X$.
When combining with the $B$-meson decay constraints, these two theoretical conditions lead to even more severe bounds for $g_X$ and $y_\mu$ than those expected at the E989 experiment.
As the result, we have
$2.53 \lesssim g_X \lesssim \sqrt{4\pi}$, and
$2.53 \lesssim |y_\mu| \lesssim \sqrt{4\pi}$.

\begin{figure}[h]
\center
\includegraphics[scale=0.34]{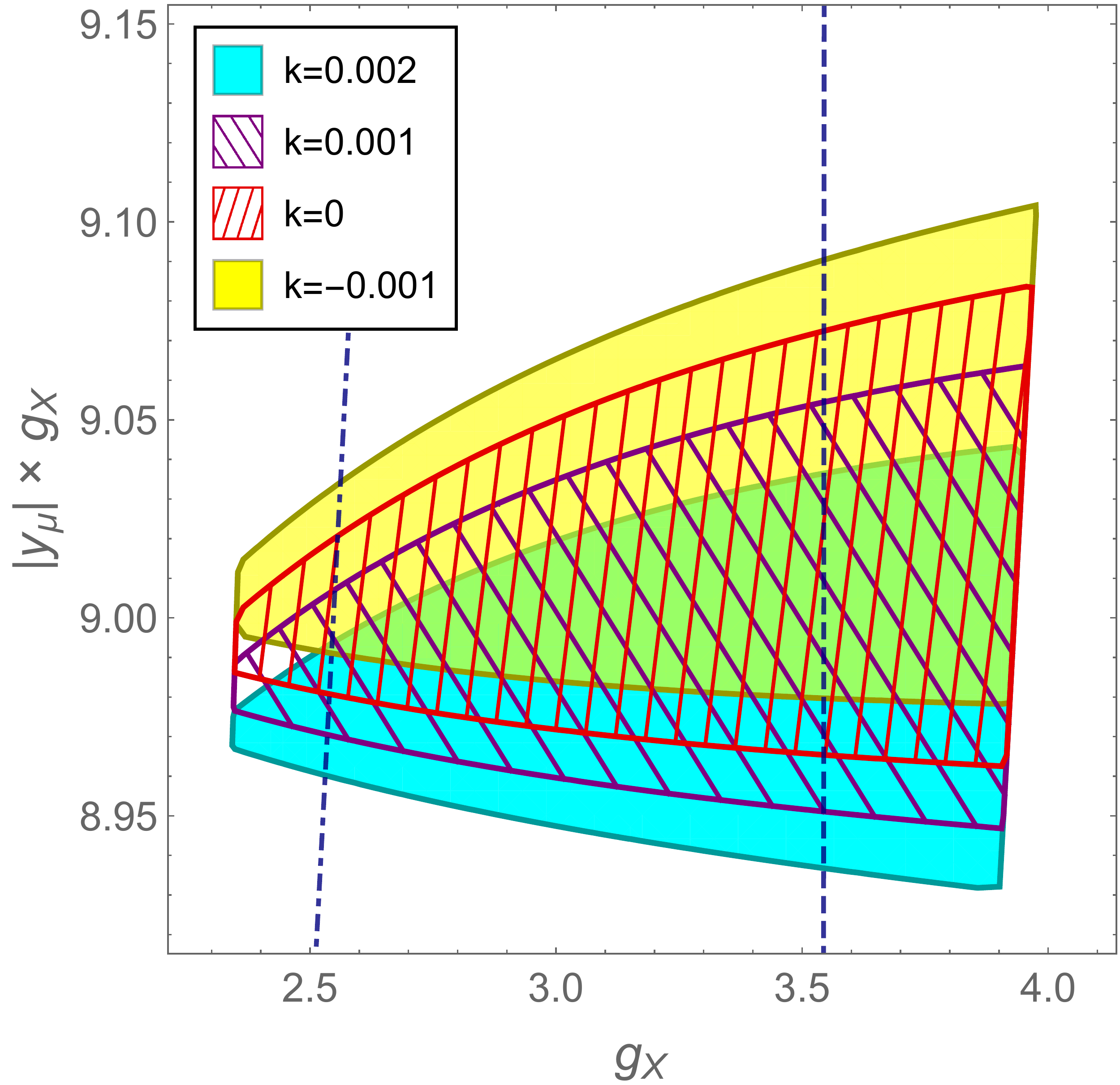}
\caption{
Viable parameter regions on the $(g_X,y_\mu g_X)$ plane
for the case of 
$m_{\chi_r}=120$ GeV, 
$m_{Z'}=300$ GeV, 
$A_{bs}=24.2 \times 10^{-5}$, 
$B_{bs}=-11.5 \times 10^{-5}$, 
$\tau=1.78$, 
$\delta=1$, 
and various values of the kinetic mixing coefficient  $k=-0.001, 0, 0.001, 0.002$.
The dot-dashed and dashed lines correspond to the perturbation limits on $y_\mu$ and $g_X$, respectively.
}
\label{gy_k_overlap}
\end{figure}

For the case of nonzero kinetic mixing,
since the effect of the kinetic mixing on the muon $g-2$ is small, the allowed range determined from the $\Delta a_\mu^\text{NP}$ constraint remains almost intact.
Although the constrained regions by the $B$-meson decay processes (the green, the yellow, the blue hatched and the red hatched regions) still have the hyperbolic shape as before, they slightly shift downward to the region with smaller values of $|y_\mu|$ (while the values of $g_X$ are fixed) in comparison to the case of vanishing kinetic mixing.
This is due to the additional contributions of the diagrams with the $Z'$-boson exchange at the tree level to the Wilson coefficients when $k \neq 0$.
As a result, the above thin strip of allowed parameter region in Figure \ref{gy_k0_proj} slightly changes with respect to the change of the kinetic mixing coefficient.
To magnify this behavior, in Figure \ref{gy_k_overlap},
we 
plot the viable parameter regions in the $(g_X, y_\mu g_X)$ plane with four benchmark values of the kinetic mixing coefficient
$k = -0.001, 0, 0.001$, and 0.002.
It is observed that,
when increasing $k$, the viable parameter region shift downward indicating that smaller values of the product $|y_\mu| g_X$ are preferable for larger values of $k$.
In other words, for a given $U(1)_X$ gauge coupling, $g_X$, smaller values of $|y_\mu|$ are more favored for larger values of $k$.
In this figure, the perturbation conditions on $y_\mu$ and $g_X$ are represented by the dot-dashed and dashed lines, respectively.
The parameter region satisfying these conditions stays in between these two straight lines.

\textbf{Constraints on the 
$(g_X, m_{Z'})$ plane:}

\begin{figure}[h]
\begin{minipage}{0.48\textwidth}
\begin{flushleft}
\includegraphics[scale=0.58]{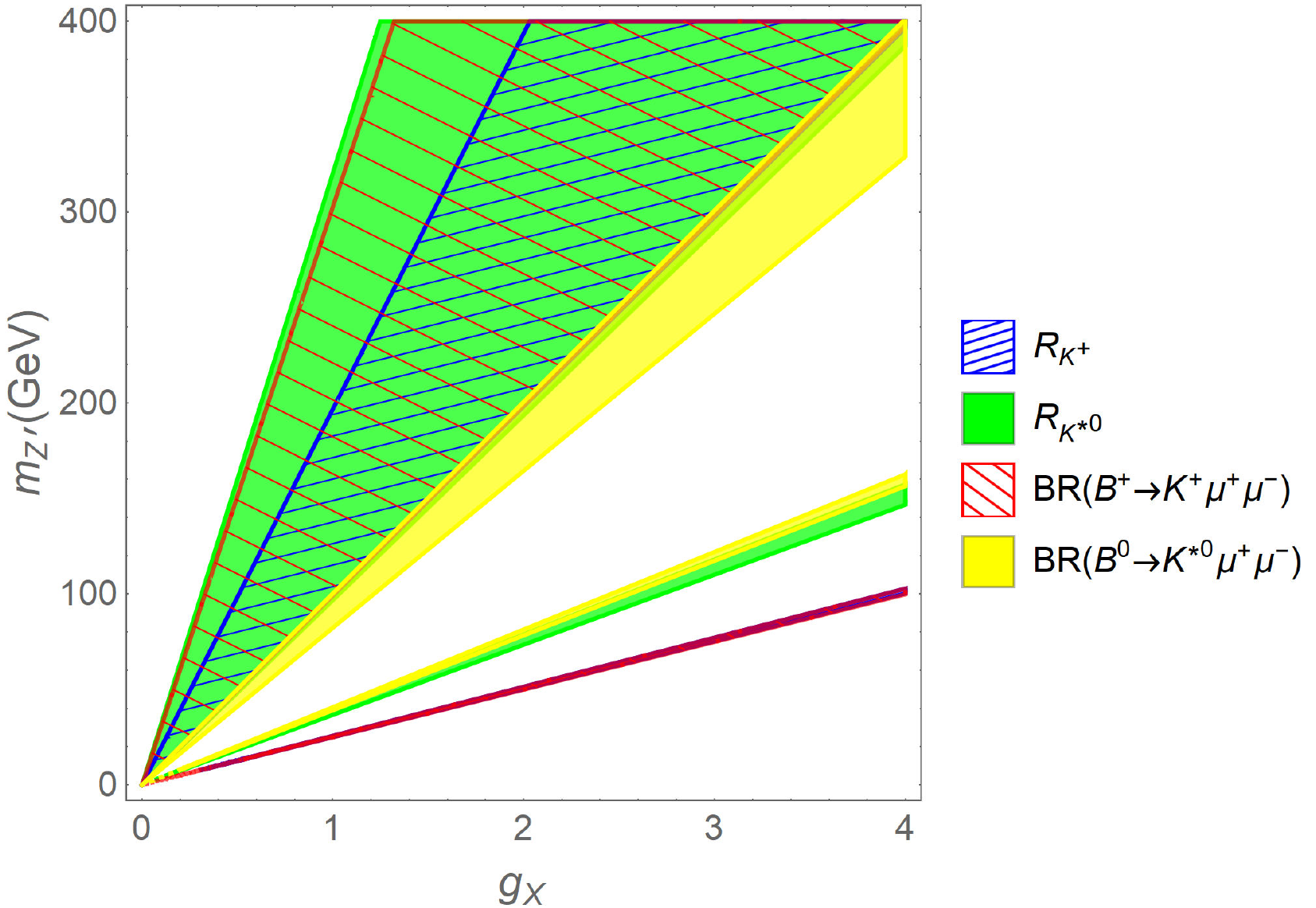}
\caption{
Phenomenological constraints on the $(g_X,m_{Z'})$ plane
for the case of 
$y_\mu = 3$, 
$A_{bs}=24.2 \times 10^{-5}$, 
$B_{bs}=-11.5 \times 10^{-5}$, 
$\tau=1.78$, 
$\delta=1$, 
and $k=0$.
}
\label{gm_k0}
\end{flushleft}
\end{minipage}
\hspace{0.5cm}
%
\begin{minipage}{0.48\textwidth}
\begin{flushright}
\includegraphics[scale=0.37]{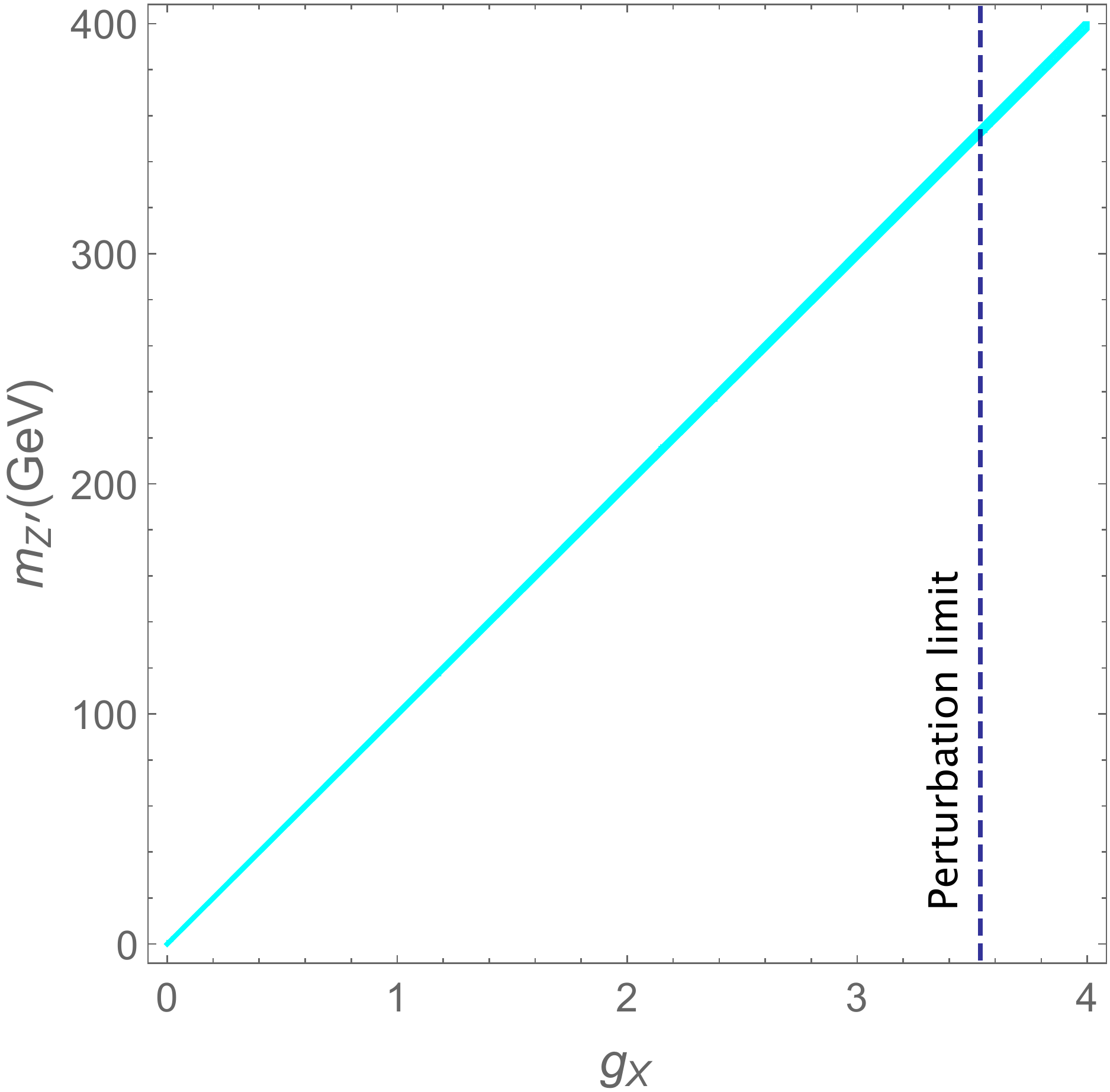}
\caption{
Viable parameter region on the $(g_X,m_{Z'})$ plane
for the case of 
$y_\mu = 3$, 
$A_{bs}=24.2 \times 10^{-5}$, 
$B_{bs}=-11.5 \times 10^{-5}$, 
$\tau=1.78$, 
$\delta=1$, 
and $k=0$.
}
\label{gm_k0_viable}
\end{flushright}
\end{minipage}
\end{figure}

The phenomenological constraints on the $(g_X, m_{Z'})$ plane are presented in Figure \ref{gm_k0} for the case of vanishing kinetic mixing and fixed values of other parameters as
$y_\mu = 3$, 
$A_{bs}=24.2 \times 10^{-5}$, 
$B_{bs}=-11.5 \times 10^{-5}$, 
$\tau=1.78$, 
and
$\delta=1$. 
For given values of the branching ratios, the parameters $g_X$ and $m_{Z'}$ are linearly dependent.
This is because these two parameters always come in the term of $\frac{g_X^2}{m_{Z'}^2}$ in
the expression of the Wilson coefficients when $k=0$.
For each constraints on $R_K$, $R_{K^*}$, $BR(B^+ \rightarrow K^+ \mu^+\mu^-)$, and
$BR(B^0 \rightarrow K^{*0} \mu^+\mu^-)$, there are two separated range of the ratio $\frac{g_X^2}{m_{Z'}^2}$.
However, there is only one overlapping region satisfying all of these four constraints.
Furthermore, this allowed region is much more severe than the overlapping region determined by only two constraints from $R_K$ and $R_{K^*}$ (the blue hatched and the green regions).
Therefore, additional consideration of the branching ratios of the decay processes
$B^+ \rightarrow K^+ \mu^+\mu^-$ and $B^0 \rightarrow K^{*0} \mu^+\mu^-$ (the red hatched and the yellow regions)
is crucial.
The viable parameter region from all of these four constraints is plotted separately in Figure \ref{gm_k0_viable}.
From this, we can determine the
allowed range for the ratio 
$\frac{m_{Z'}}{g_X}$ to be 
98.2$-$100.8 GeV.
Taking into account the perturbation limit for $g_X$, we find the upper bound for $m_{Z'}$ to be approximately 
354 GeV in this case.

\begin{figure}[h]
\begin{minipage}{0.48\textwidth}
\begin{flushleft}
\includegraphics[scale=0.56]{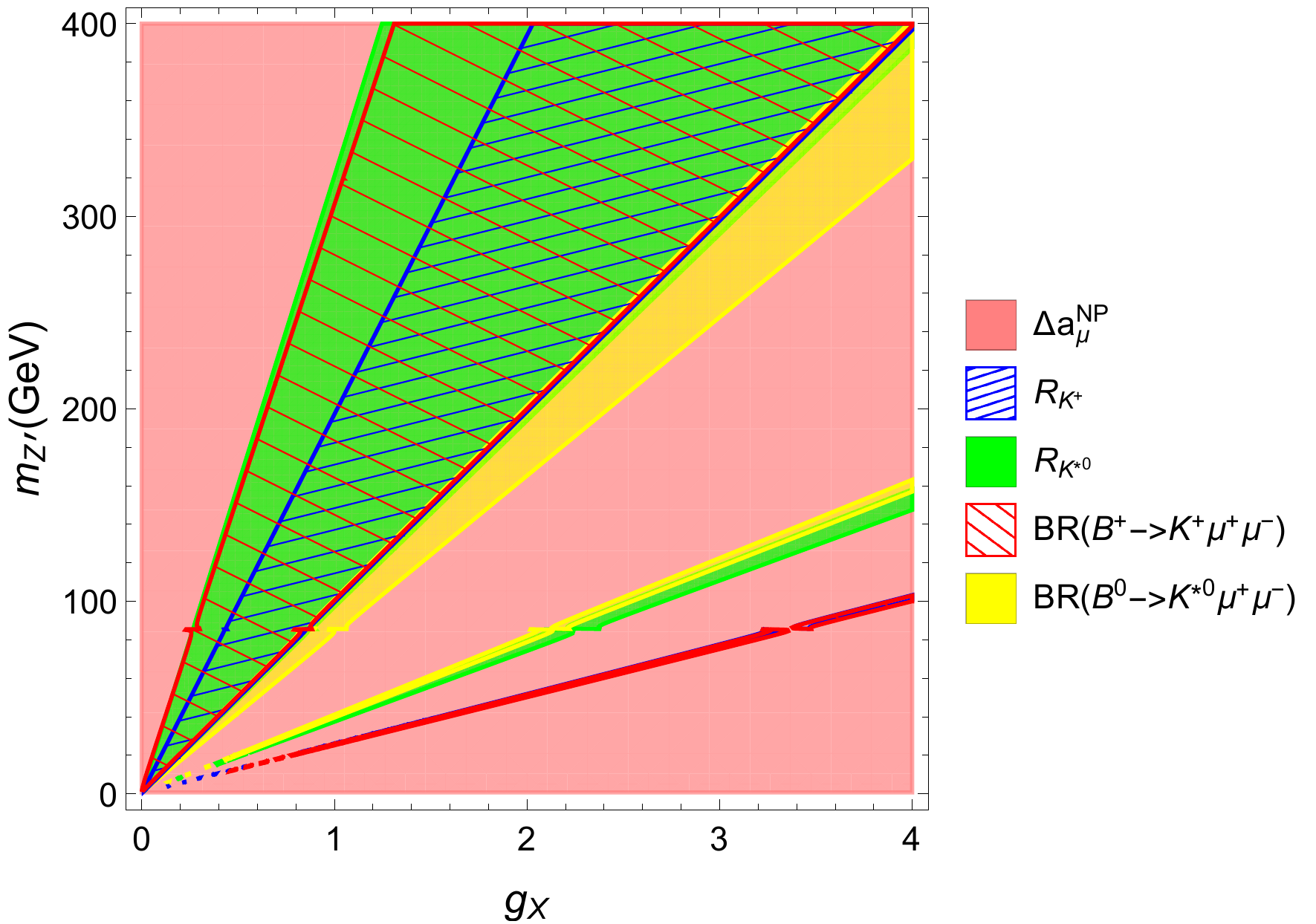}
\caption{
Phenomenological constraints on the $(g_X,m_{Z'})$ plane
for the case of 
$m_{\chi_r} = 120$ GeV,
$y_\mu = 3$, 
$A_{bs}=24.2 \times 10^{-5}$, 
$B_{bs}=-11.5 \times 10^{-5}$, 
$\tau=1.78$, 
$\delta=1$, 
and $k=0.002$.
}
\label{gm_k0.002}
\end{flushleft}
\end{minipage}
\hspace{0.5cm}
%
\begin{minipage}{0.48\textwidth}
\begin{flushright}
\includegraphics[scale=0.2]{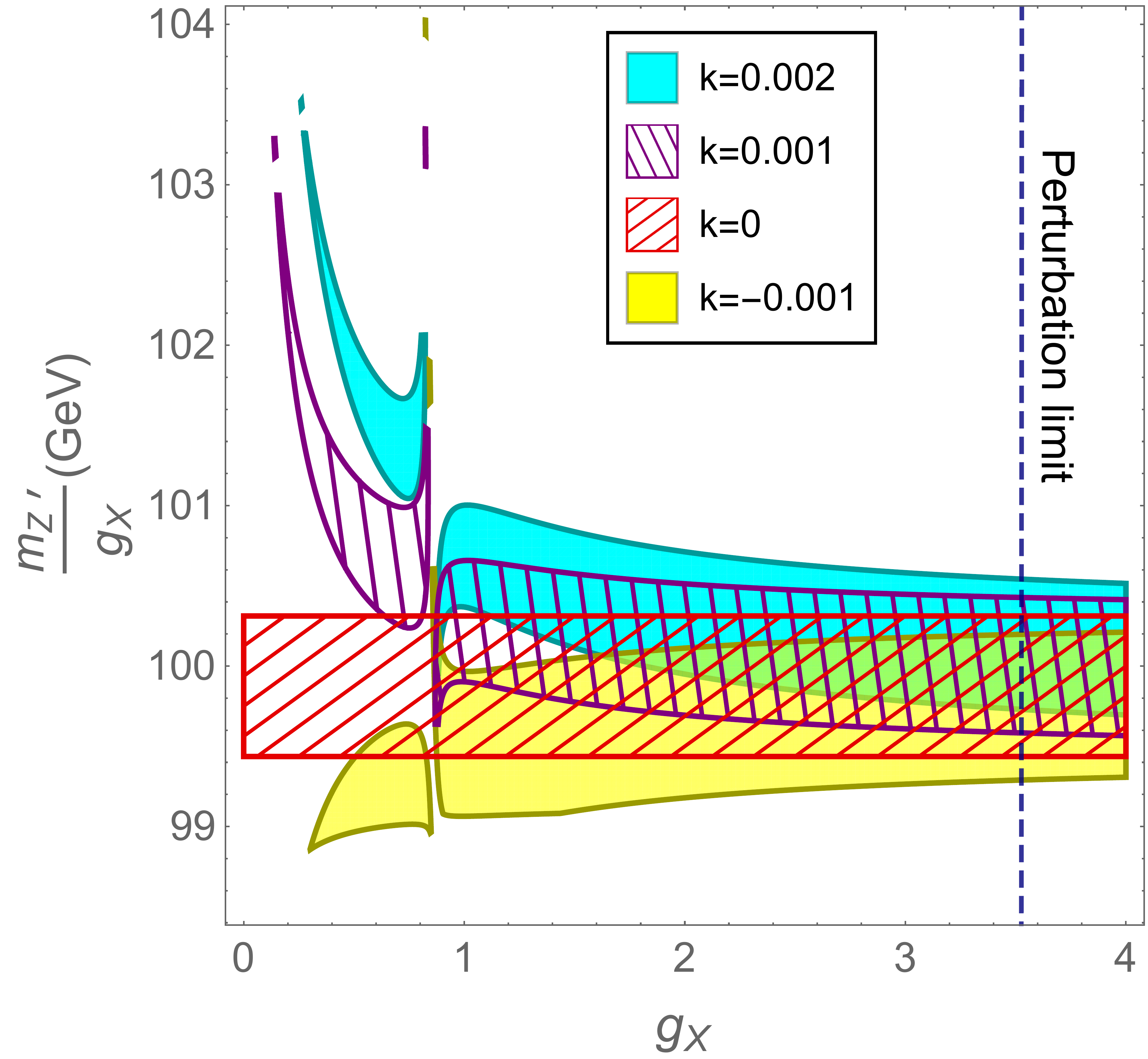}
\caption{
Viable parameter regions on the $(g_X,\frac{m_{Z'}}{g_X})$ plane
for the case of 
$m_{\chi_r} = 120$ GeV,
$y_\mu = 3$, 
$A_{bs}=24.2 \times 10^{-5}$, 
$B_{bs}=-11.5 \times 10^{-5}$, 
$\tau=1.78$, 
$\delta=1$, 
and various values of the kinetic mixing coefficient  $k=-0.001, 0, 0.001, 0.002$.
}
\label{gm_k_overlap}
\end{flushright}
\end{minipage}
\end{figure}

When the gauge kinetic mixing is switched on, the linear correlation between $g_X$ and $m_{Z'}$ is deformed due to nonzero values of $k$ in the Wilson coefficients.
In Figure \ref{gm_k0.002}, we demonstrate the phenomenological constraints on the $(g_X, m_{Z'})$ plane for $k=0.002$.
Here, the deformation appears when $m_{Z'} \approx 85$ GeV is due to the sign flipping of the $ZZ'$ mixing angle, $\alpha_Z$, in Eq. (\ref{alphaZ}).
%
%
%
%
In the case with nonzero gauge kinetic mixing, the constraint from the muon $g-2$ measurement need to be considered since $\Delta a_\mu^\text{NP}$ depends on $m_{Z'}$ via the $\beta$ term in Eq. (\ref{amu_k}).
For $k=0.002$, we find the lower bound for the $Z'$-boson mass from this constraint to be 
$m_{Z'} \gtrsim 0.68$ GeV.
In Figure \ref{gm_k_overlap}, we plot the viable parameter regions for various values of the kinetic mixing coefficients 
($k=-0.001, 0, 0.001$, and 0.002)
in the $(g_X, \frac{m_{Z'}}{g_X})$ plane.
Comparing these areas, we see that, for larger values of $k$,
the allowed region 
in this plane shifts upward implying that
larger values of $m_{Z'}$ are more favored for a given value of $g_X$.
Similar to Figure \ref{gm_k0_viable}, the perturbation limit for $g_X$ is important in ruling out the region with large $Z'$-boson mass. It sets the upper limit for $m_{Z'}$ for fixed values of other parameters.
It is seen that larger values of the kinetic mixing coefficient $k$ lead to slightly larger upper bounds on the $Z'$-boson mass.

\textbf{Constraints on the 
$(y_\mu, m_{Z'})$ plane:}

\begin{figure}[h]
\begin{minipage}{0.48\textwidth}
\begin{flushleft}
\includegraphics[scale=0.56]{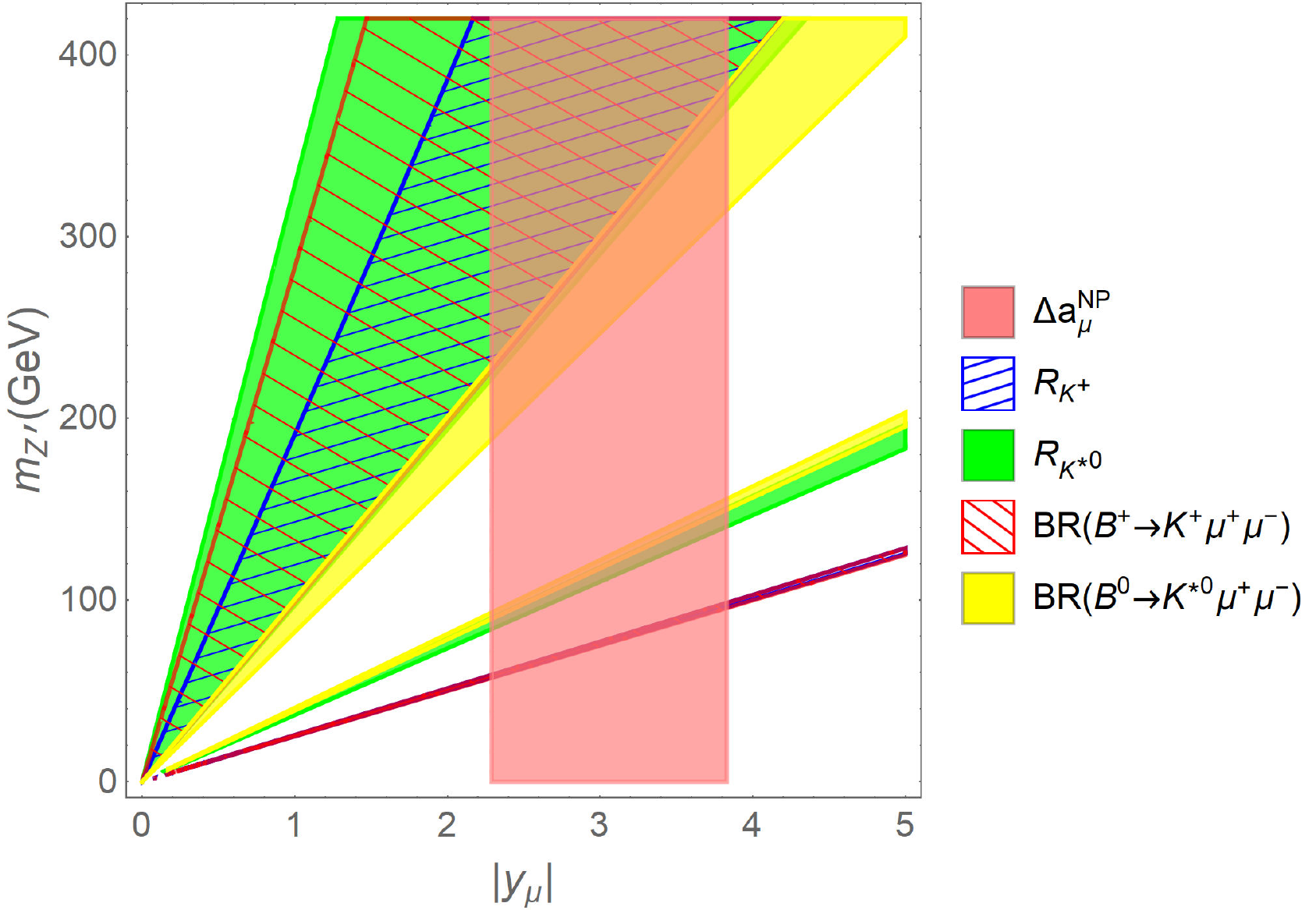}
\caption{
Phenomenological constraints on the $(y_\mu,m_{Z'})$ plane
for the case of 
$m_{\chi_r}=120$ GeV, 
$g_X = 3$, 
$A_{bs}=24.2 \times 10^{-5}$, 
$B_{bs}=-11.5 \times 10^{-5}$, 
$\tau=1.78$, 
$\delta=1$, 
and $k=0$.
}
\label{ym_k0}
\end{flushleft}
\end{minipage}
\hspace{0.5cm}
%
\begin{minipage}{0.48\textwidth}
\begin{flushright}
\includegraphics[scale=0.38]{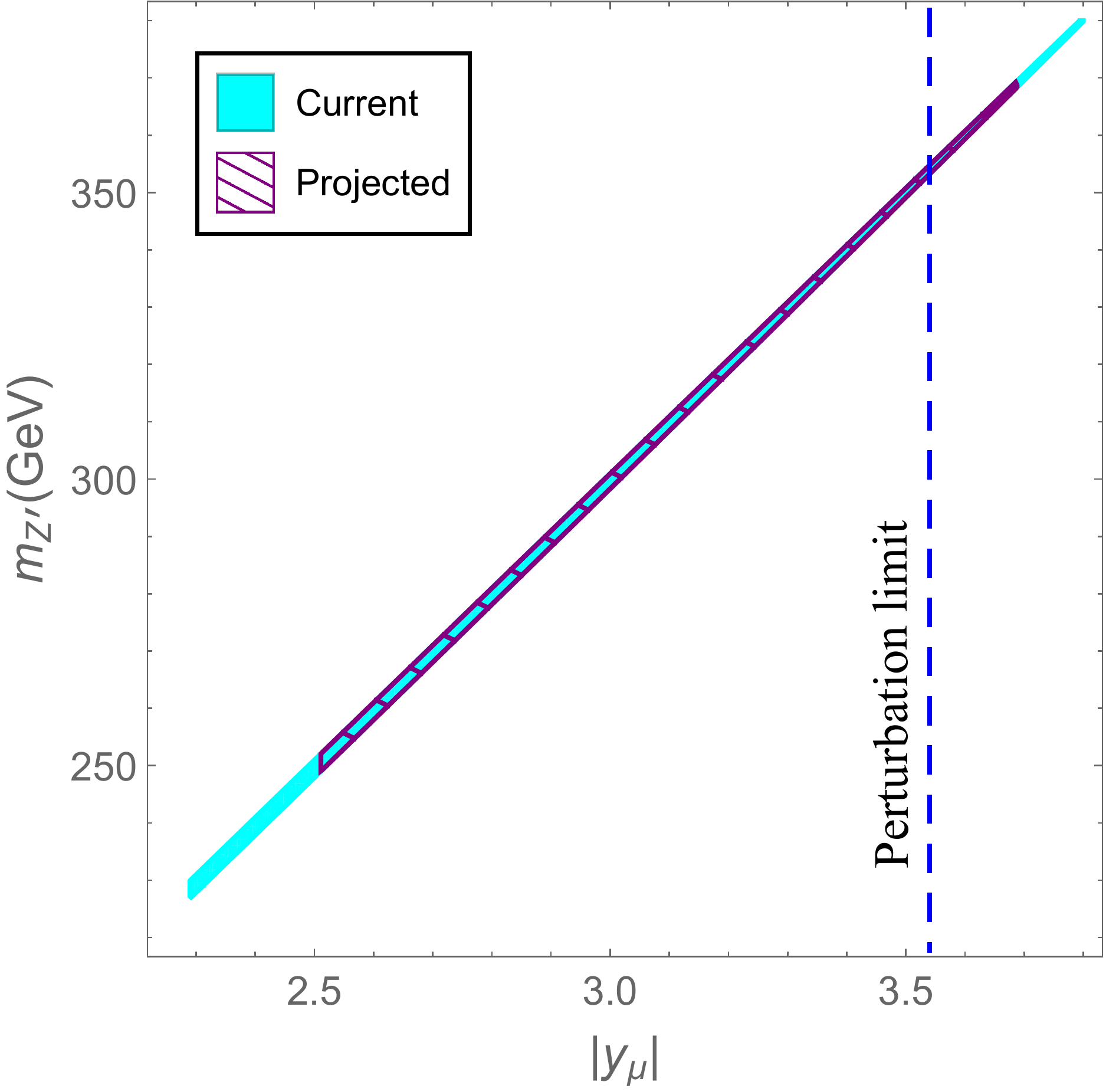}
\caption{
Viable parameter region on the $(y_\mu,m_{Z'})$ plane
for the case of 
$m_{\chi_r}=120$ GeV, 
$g_X = 3$, 
$A_{bs}=24.2 \times 10^{-5}$, 
$B_{bs}=-11.5 \times 10^{-5}$, 
$\tau=1.78$, 
$\delta=1$,
and $k=0$.
}
\label{ym_k0_pro}
\end{flushright}
\end{minipage}
\end{figure}

The phenomenological constraints on the $(y_\mu, m_{Z'})$ plane are plotted in Figure \ref{ym_k0} where the values of other fixed inputs are 
$m_{\chi_r}=120$ GeV, 
$g_X = 3$, 
$A_{bs}=24.2 \times 10^{-5}$, 
$B_{bs}=-11.5 \times 10^{-5}$, 
$\tau=1.78$, 
$\delta=1$,
and $k=0$, respectively.
Since $m_{Z'}$ is not involved in the new physics contribution at the one-loop level to the muon $g-2$ when $k=0$, the constraint on $\Delta a_\mu^\text{NP}$ is a vertical band in this figure.
This constraint leads to the same limits for $y_\mu$ as those in Figure \ref{gy_k0}.
Similar to Figure \ref{gm_k0}, the $B$-meson decay width leads to an 
approximately linear dependence between these two parameters ($y_\mu$ and $m_{Z'}$) on the plane due to the factor $\frac{|y_\mu|}{m_{Z'}}$ in the Wilson coefficients when there is no kinetic mixing.
In particular, each of the constraints on $R_K$, $R_{K^*}$, $BR(B^+ \rightarrow K^+ \mu^+\mu^-)$, and 
$BR(B^0 \rightarrow K^{*0} \mu^+\mu^-)$ determines two allowed ranges for the values of the ratio $\frac{|y_\mu|}{m_{Z'}}$.
However, there is only one overlapping region 
that satisfies all of the considered 2$\sigma$ bounds.
This region is shown separately in Figure \ref{ym_k0_pro} with the cyan color.
Since the allowed region is a thin strip, we find that the correlation among the two considered parameters is
$\frac{m_{Z'}}{g_X} \sim 100$ GeV for the given set of other inputs.
From this, we can extract the limits for the $Z'$-boson mass at 95\% C.L. as
$226 \text{ GeV} 
\lesssim m_{Z'} \lesssim 381 \text{ GeV}$.
In the near future, when the E989 experiment get the full data, assuming the center value of the muon $g-2$ remains the same, we can expect the new limits for the $Z'$-boson mass to narrow down to 
$248 \text{ GeV} \lesssim m_{Z'} \lesssim 370 \text{ GeV}$
as illustrated by the hatched region.
Taking into account the perturbation condition for $y_\mu$, the upper bound for $m_{Z'}$ is even more severe than the one imposed by the projected E989 result, namely, 
$m_{Z'} \lesssim 355$ GeV.

\begin{figure}[h!]
\begin{minipage}{0.48\textwidth}
\begin{flushleft}
\includegraphics[scale=0.56]{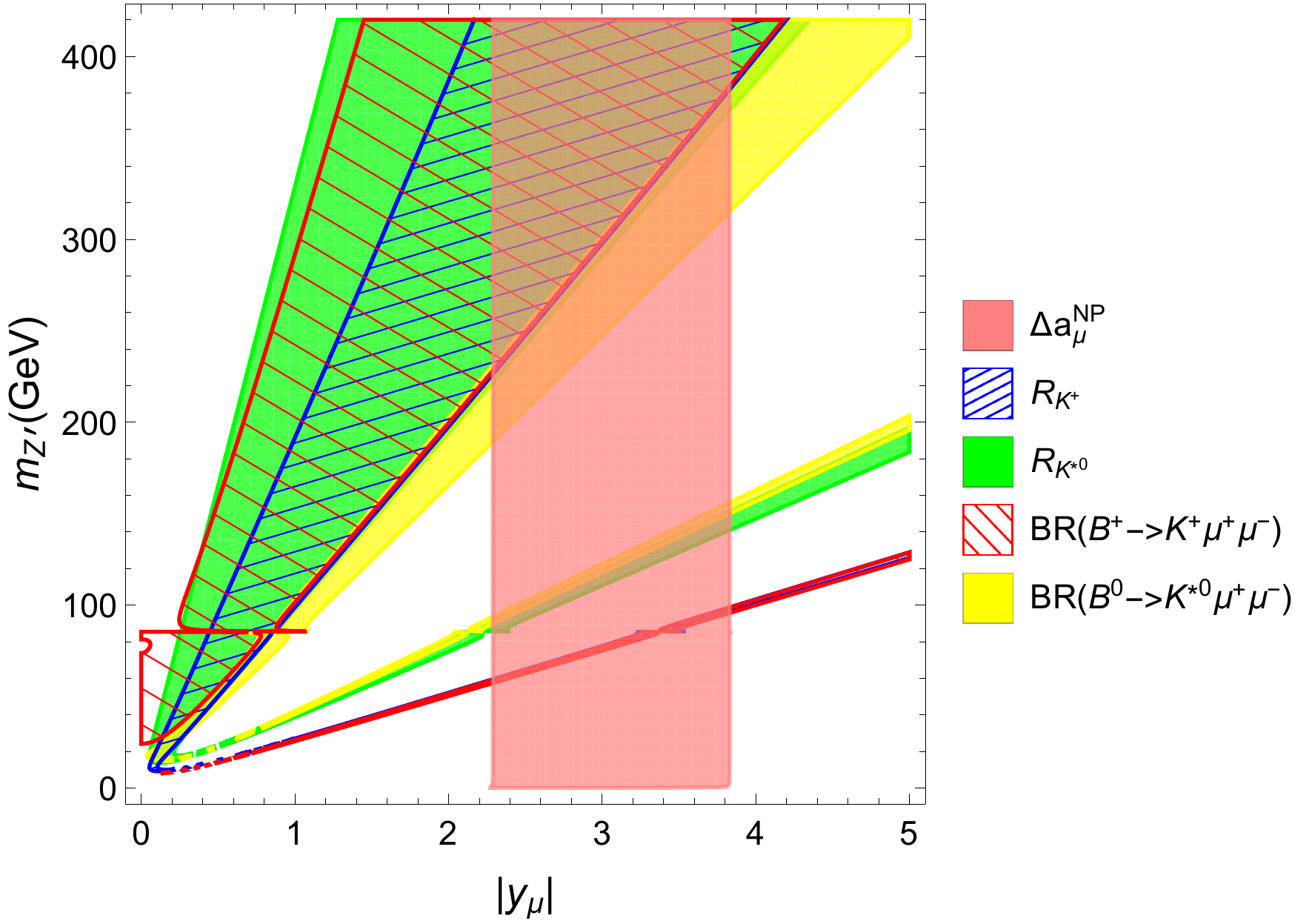}
\caption{
Phenomenological constraints on the $(y_\mu,m_{Z'})$ plane
for the case of 
$m_{\chi_r}=120$ GeV, 
$g_X = 3$, 
$A_{bs}=24.2 \times 10^{-5}$, 
$B_{bs}=-11.5 \times 10^{-5}$, 
$\tau=1.78$, 
$\delta=1$,
and $k=0.002$.
}
\label{ym_k0.002}
\end{flushleft}
\end{minipage}
\hspace{0.5cm}
%
\begin{minipage}{0.48\textwidth}
\begin{flushright}
\includegraphics[scale=0.42]{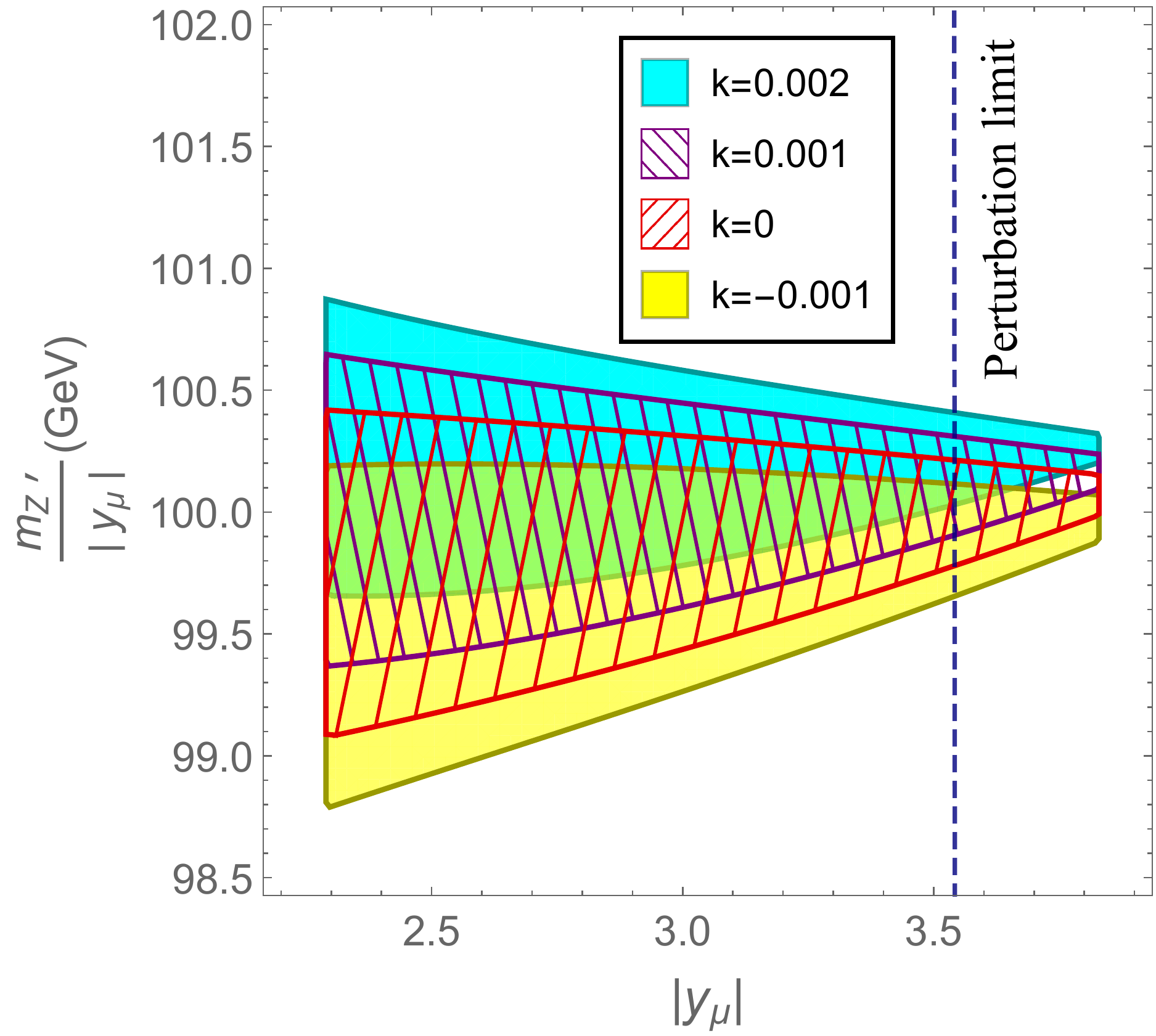}
\caption{
Viable parameter regions on the $(y_\mu,m_{Z'})$ plane
for the case of 
$m_{\chi_r}=120$ GeV, 
$g_X = 3$, 
$A_{bs}=24.2 \times 10^{-5}$, 
$B_{bs}=-11.5 \times 10^{-5}$, 
$\tau=1.78$, 
$\delta=1$,
and various values of the kinetic mixing coefficient  $k=-0.001, 0, 0.001, 0.002$.
}
\label{ym_k_overlap}
\end{flushright}
\end{minipage}
\end{figure}

When the gauge kinetic mixing is introduced, the constrained regions are strongly distorted as depicted in Figure \ref{ym_k0.002} for the case of $k=0.002$.
On the one hand, regarding the constraint on 
$\Delta a_\mu^\text{NP}$ (the pink region) 
we see that the new Yukawa coupling, $y_\mu$, can be small in the presence of the kinetic mixing term as the $Z'$ boson is light enough with its mass of less than 
$\mathcal{O}$(1) GeV.
On the other hand, the constraint on the branching ratio of the decay process
$B^+ \rightarrow K^+ \mu^+\mu^-$ imply that $m_{Z'}$ must be larger than about 
8 GeV.
Therefore, small values of $|y_\mu|$ are forbidden in the case $k=0.002$.
The blue hatched regions (the green regions,
the yellow regions) 
determined by the bounds on $R_K$ ($R_{K^*}$,  
$BR(B^0 \rightarrow K^{*0} \mu^+ \mu^-)$)
which contain two separated ranges of the ratio $\frac{|y_\mu|}{m_{Z'}}$ as in Figure \ref{ym_k0}, become connected in this case. 
The red hatched regions determined by the bounds on $BR(B^+ \rightarrow K^+ \mu^+\mu^-)$ also experience strong deformations particularly when $8 \text{ GeV} \lesssim m_{Z'} \lesssim 90 \text{ GeV}$.
Especially, all the constrained regions are deformed when the $Z'$-boson mass is close to about 85 GeV where the mixing angle $\alpha_Z$ changes its sign.
The viable parameter regions on the plane $(y_\mu, \frac{m_{Z'}}{y_\mu} )$ satisfying all the considered constraints are presented in Figure \ref{ym_k_overlap} for different values of the gauge kinetic mixing coefficient 
$k=-0.001$, 0, 0.001, and 0.002.
We observe that the viable region shifts up when increasing the coefficient $k$.
This indicates that larger values of $m_{Z'}$ are more favored for larger values of $k$ as $y_\mu$ is fixed.
Meanwhile, the allowed 
range for $y_\mu$ is almost independent on the gauge kinetic mixing coefficient $k$.

\textbf{Constraints on $k$:}

\begin{figure}[h]
\begin{center}
\includegraphics[scale=0.5]{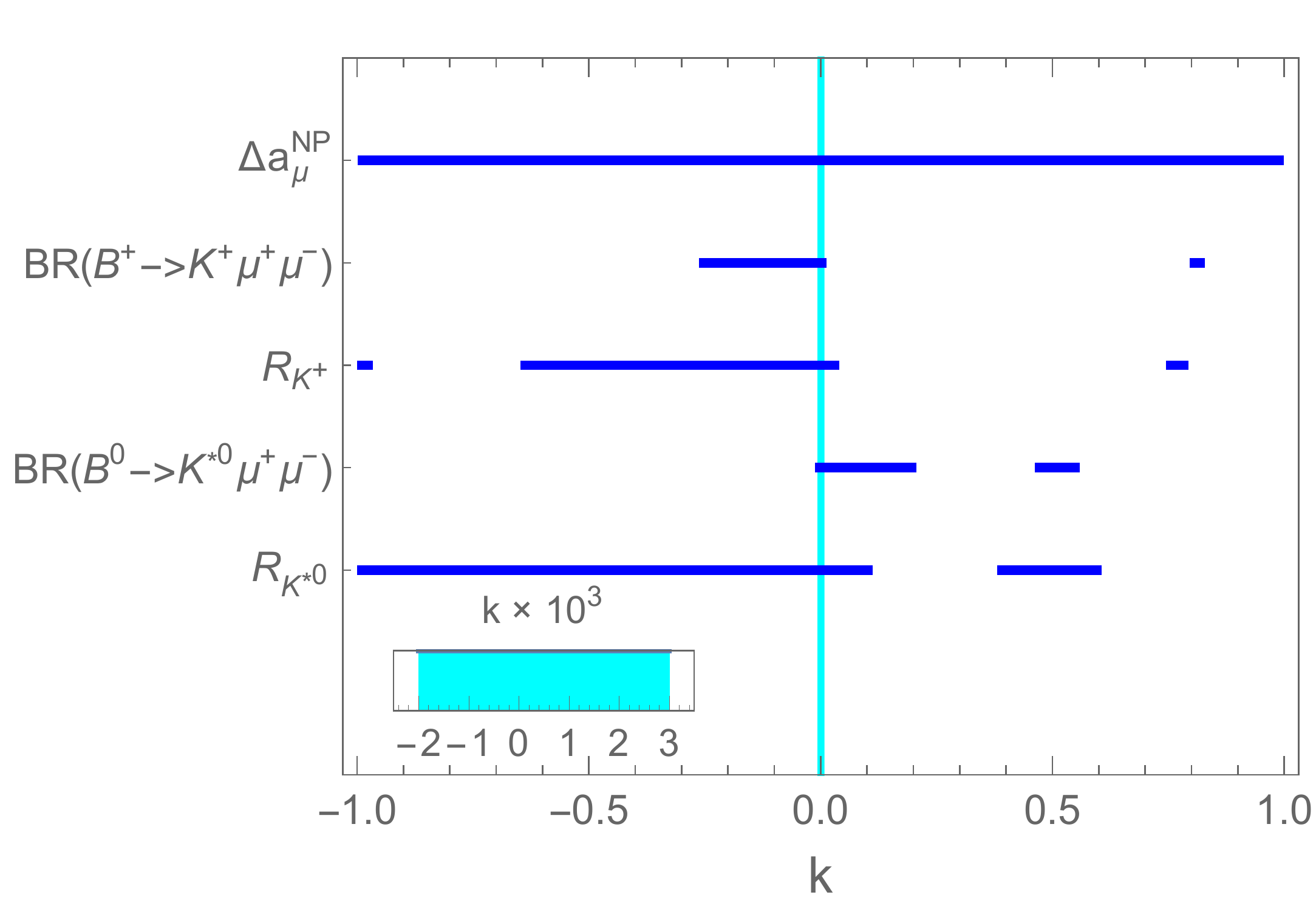}
\caption{
Phenomenological constraints on the gauge kinetic mixing coefficient
for the case of 
$m_{\chi_r}=120$ GeV, 
$m_{Z'}=300$ GeV 
$g_X = 3$, 
$y_\mu = 3$, 
$A_{bs}=24.2 \times 10^{-5}$, 
$B_{bs}=-11.5 \times 10^{-5}$, 
$\tau=1.78$, 
$\delta=1$. 
}
\label{constraints_on_k}
\end{center}
\end{figure}

The phenomenological constraints on the gauge kinetic mixing coefficient $k$ are plotted in Figure \ref{constraints_on_k} when other parameters are fixed as
$m_{\chi_r}=120$ GeV, 
$m_{Z'}=300$ GeV 
$g_X = 3$, 
$y_\mu = 3$, 
$A_{bs}=24.2 \times 10^{-5}$, 
$B_{bs}=-11.5 \times 10^{-5}$, 
$\tau=1.78$, 
and $\delta=1$.
We observe that the constraint on $\Delta a_\mu^\text{NP}$ only excludes large values of $k$ being close to $\pm 1$.
The $R_K$ and $R_{K^*}$ constraints are quite severe since they rule out large portions of the possible range of $k$.
The most stringent limits on $k$ are given by the constraints on the branching ratios of the decay processes 
$B^+ \rightarrow K^+ \mu^+ \mu^-$ and 
$B^0 \rightarrow K^{*0} \mu^+ \mu^-$.
There are two narrow ranges of $k$ satisfying each of these two constraints.
This observation once again reveals how important these branching ratios are beside $R_K$ and $R_{K^*}$.
By overlapping all the ranges of $k$ allowed by the considered constraints, we find the viable range of the kinetic mixing coefficient to be 
$-0.002 \lesssim k \lesssim 0.003$
 when other parameters are fixed.
This range is magnified in the figure by its inset.
This result shows that the upper bound for the kinetic mixing coefficient in the case $m_{Z'}= 300$ GeV is comparable with that derived from other experimental data in the case $m_{Z'} \lesssim 200$ GeV as in Eq. (\ref{upper_bound_k}). 
Therefore, the $B$-meson decays are important channels to constraint the parameter space.
At the one-loop level, the kinetic mixing coefficient is estimated \cite{Cheung:2009qd} to be at the magnitude of about
$\mathcal{O}(10^{-1})-\mathcal{O}(10^{-2})$
that is already larger than the current upper bound.
Therefore, in this case, $k$ must be nonzero at the tree level with a similar order of magnitude and an opposite sign so that the total effective value of this parameter is consistent with the experimental bounds.

In Table \ref{benchmark}, as a demonstration, the relevant observables are calculated for four benchmark values of the gauge kinetic mixing 
$k=-0.002$, 0, 0.002, and 0.003 while other parameters are chosen the same as those in Figure \ref{constraints_on_k}.
All of them satisfy the corresponding experimental bounds at the $2\sigma$ level.
We notice that, in this table, while the values of $\Delta a_\mu^\text{NP}$ only show the differences from the eighth significant digits when changing $k$, differences for the values of other observables can be seen from the third or the forth significant digits.
It implies that the gauge kinetic mixing coefficient has a much smaller effect on the muon $g-2$ than those on the semileptonic $B$ decays and the violation of lepton universality.
The FCNC parameters in the model, $A_{bs}$ and $B_{bs}$, induce the tree-level contributions to the $B_s^0-\bar{B}_s^0$ mixing.
Due to the constraints on the rare $B$-meson decays, these parameters are required to be very small of about $\mathcal{O}(10^{-4})$.
Therefore, such contributions to the mixing observables, $\Delta m_s$ and $\Delta \Gamma_s$, are negligible in comparison to the SM contributions%
\footnote{Note that the theoretical uncertainties of the SM predictions for $\Delta m_s$ and $\Delta \Gamma_s$ are respectively about 5\% and 23\% that are relatively large
\cite{Bona:2006ah,Zyla:2020zbs}.
} 
that are consistent with the experimental values \cite{Zyla:2020zbs}.
Beside the mixing of vectorlike and SM quarks,
the gauge kinetic mixing induces additional tree-level contributions to the couplings between the $Z'$ boson and all SM flavors that are approximately proportional to $k \lesssim \mathcal{O}(10^{-3})$. 
Therefore, the $Z'$ production cross sections at hadron colliders, as well as its decay widths are slightly modified by small amounts roughly proportional to $k^2 \lesssim \mathcal{O}(10^{-6})$ in comparison to the case of vanishing kinetic mixing.
Assuming the existence of the kinetic mixing, we have calculated the $Z'$ production cross section times branching ratio to dimuon of $pp$ collisions at $\sqrt{s}=13$ TeV.
It is found to be about $\mathcal{O}(1)$ fb 
that is of the same order as the experimental limit in Ref. \cite{ATLAS:2019erb}.
Thus, the benchmark point is marginally acceptable.
In the near future, more precise analyses at the LHC will be able to test the model.

\begin{table}[h]
\begin{center}
\begin{math}
\begin{array}{|c||c|c|c|c|c|}
\hline
k& \Delta a_{\mu}^{\text{NP}}& BR (B^+ \rightarrow K^+ \mu^+ \mu^-) & R_K & BR(B^0 \rightarrow K^{*0} \mu^+ \mu^-) & R_{K^*} \\
\hline 
\hline
 -0.002 & 2.2698174\times 10^{-9} & 1.05535\times 10^{-7} & 0.76795 & 1.96886\times 10^{-7} & 0.55552 \\
 0 & 2.2698173\times 10^{-9} & 1.05332\times 10^{-7} & 0.76772 & 1.96171\times 10^{-7} & 0.55505 \\
 0.002 & 2.2698174\times 10^{-9} & 1.05129\times 10^{-7} & 0.76749 & 1.95459\times 10^{-7} & 0.55458 \\
 0.003 & 2.2698176\times 10^{-9} & 1.05028\times 10^{-7} & 0.76737 & 1.95104\times 10^{-7} & 0.55434 \\
\hline
\end{array}
\end{math}
\caption{The considered observables
for the case of 
$m_{\chi_r}=120$ GeV, 
$m_{Z'}=300$ GeV 
$g_X = 3$, 
$y_\mu = 3$, 
$A_{bs}=24.2 \times 10^{-5}$, 
$B_{bs}=-11.5 \times 10^{-5}$, 
$\tau=1.78$, 
$\delta=1$,
 and four benchmark values of the gauge kinetic mixing coefficient $k$.}
\label{benchmark}
\end{center}
\end{table}


The particle $\chi_r$ in this model is stable and neutral under the SM gauge groups. 
It can be a candidate for dark matter.
In the original BDW model, the leading contribution to the spin-independent cross section between $\chi_r$ and nucleon was estimated to be $\sigma_{SI}^p \sim \mathcal{O}(10^{-50})$ cm$^2$
\cite{Belanger:2015nma}.
In our analysis, the nonzero kinetic mixing slightly enhances the chance of the spin-independent scattering by allowing additional one-loop contributions.
However, since the kinetic mixing is limited to be $k \sim \mathcal{O}(10^{-3})$, such new one-loop contributions to the spin-independent cross section is suppressed.
As a result, the total cross section is still smaller than the coherent neutrino-nucleus scattering background \cite{Billard:2013qya}.
With the chosen parameter sets, the pair annihilation process of $\chi_r$ into a pair of leptons ($\mu$ or $\nu_\mu$) is effective 
due to the large $y_\mu$ coupling.
In addition, the coannihilation between the vectorlike lepton and $\chi_r$ also reduces its relic density since their masses are relatively close.
Therefore, the relic density of $\chi_r$ is smaller than the observed dark matter abundance as obtained in Ref. \cite{Belanger:2015nma}.
Once the kinetic mixing is switched on, the above pair annihilation and coannihilation processes happen more frequently in the early universe leading to a smaller value of $\Omega_{\chi_r} h^2$ than that in the case of vanishing kinetic mixing.
In order to account for the dark matter relic density measured by the Planck Collaboration \cite{Aghanim:2018eyx},
an additional dark matter candidate is necessary.

\section{Conclusion}

The BDW model with additional vectorlike lepton and quark doublets and two complex scalars charged under the $U(1)_X$ gauge group
is well motivated due to its ability in explaining various anomalies at the same time.
In this paper, we have generalized this model by introducing the gauge kinetic mixing term.
The new physics contributions to the muon anomalous magnetic moment and the Wilson coefficients ($C_{9,10}^{(')}$) have been calculated analytically.
We have investigated the parameter space of the model taking into account the phenomenological constraint on the muon $g-2$, the updated LHCb data on lepton universality violation ($R_K$ and $R_{K^*}$), the branching ratios of the semileptonic rare decays 
($B^+ \rightarrow K^+ \mu^+ \mu^-$ and
$B^0 \rightarrow K^0 \mu^+ \mu^-$), 
the LEP data on slepton searhces, and the LHC 13 TeV data on both slepton and $Z'$-boson searches,
as well as requirements from the perturbative theory.
The viable parameter regions satisfying all the considered constraints at the level of $2\sigma$ have been identified.
The results indicate that the FCNC parameters, $A_{bs}$ and $B_{bs}$, are 
small enough
 to be consistent with experiment data.
In the presence of the gauge kinetic mixing term, the allowed regions are shifted and deformed in comparison to the case with $k=0$.
Especially, the kinetic mixing coefficient plays an important role in extending the viable parameter regions.
The analysis also shows that the rare $B$-meson decays are important channels that provide important constraint on the gauge kinetic mixing beside the $Z'$ resonance searches.
In the near future, with the projected sensitivity, the E989 experiment will be able to test certain parts of the free parameter space and put a more severe constraint on the acceptable parameter regions of the model.
Sine the $Z'$ production cross section times branching ratio to dimuon at the LHC is of the same order as the current limit for certain parameter regions, the model can be tested in more precise analyses of this channel.



\section*{Acknowledgment}

We would like to thank Prof. Nguyen Xuan Han for his enthusiasm and encouragement on  this work.
%



\end{document}